\documentclass[10pt,conference]{IEEEtran}
\usepackage{cite}
\usepackage{amsmath,amssymb,amsfonts}
\usepackage{algorithmic}
\usepackage{graphicx}
\usepackage{textcomp}
\usepackage{xcolor}
\usepackage[hyphens]{url}
\usepackage{fancyhdr}
\usepackage{hyperref}

\usepackage{tikz}
\usepackage{enumitem}
\usepackage{filecontents}
\usepackage{makecell}
\usepackage{multirow}
\usepackage{pifont}
\usepackage{algorithm}
\usepackage[normalem]{ulem}
\usepackage{bm}
\usepackage{diagbox}
\usepackage{xcolor}

\title{\huge HybridEP: Scaling Expert Parallelism to Cross-Datacenter Scenario via Hybrid Expert/Data Transmission}
\def\hpcacameraready{} % Uncomment to build camera-ready version

\newcommand\hpcaauthors{Weihao Yang$\dagger$, Hao Huang$\dagger$, Donglei Wu$\ddagger$, Ningke Li$\dagger$, Yanqi Pan$\dagger$, Qiyang Zheng$\dagger$, \\ Wen Xia$\dagger$, Shiyi Li$\dagger$ and Qiang Wang$\dagger$}
\newcommand\hpcaaffiliation{Harbin institute of Technology, Shenzhen$\dagger$, Guangzhou University$\ddagger$}
\author{
  \ifdefined\hpcacameraready
    \IEEEauthorblockN{\hpcaauthors{}}
      \IEEEauthorblockA{
        \hpcaaffiliation{} \\
      }
  \else
    \IEEEauthorblockN{\normalsize{HPCA \hpcayear{} Submission
      \textbf{\#\hpcasubmissionnumber{}}} \\
      \IEEEauthorblockA{
        Confidential Draft \\
        Do NOT Distribute!!
      }
    }
  \fi 
}

\fancypagestyle{camerareadyfirstpage}{%
  \fancyhead{}
  
  \fancyhead[C]{
    \ifdefined\aeopen
    \parbox[][12mm][t]{13.5cm}{\hpcayear{} IEEE International Symposium on High-Performance Computer Architecture (HPCA)}    
    \else
      \ifdefined\aereviewed
      \parbox[][12mm][t]{13.5cm}{\hpcayear{} IEEE International Symposium on High-Performance Computer Architecture (HPCA)}
      \else
      \ifdefined\aereproduced
      \parbox[][12mm][t]{13.5cm}{\hpcayear{} IEEE International Symposium on High-Performance Computer Architecture (HPCA)}
      \else
      \parbox[][0mm][t]{13.5cm}{\hpcayear{} IEEE International Symposium on High-Performance Computer Architecture (HPCA)}
    \fi 
    \fi 
    \fi 
    \ifdefined\aeopen 
      \includegraphics[width=12mm,height=12mm]{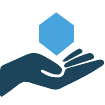}
    \fi 
    \ifdefined\aereviewed
      \includegraphics[width=12mm,height=12mm]{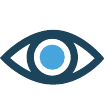}
    \fi 
    \ifdefined\aereproduced
      \includegraphics[width=12mm,height=12mm]{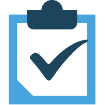}
    \fi
  }
  \fancyfoot[C]{}
}
\begin{document}
\maketitle

%Enables the camera ready header and footer
\thispagestyle{plain}
\pagestyle{plain}
% \fi

\newcommand{\hpcaheight}{0mm}
\ifdefined\eaopen
\renewcommand{\hpcaheight}{12mm}
\fi

%%%%%%%%%%%%%%%%%%%%%%%%%%%%%%%%%%%%%%%%
%%%%%%%% -- PAPER CONTENT STARTS -- %%%%%%%%%

\begin{abstract}
Mixture-of-Experts (MoE) has become a popular architecture for scaling large models.
However, the rapidly growing scale outpaces model training on a single DC, driving a shift toward a more flexible, cross-DC training paradigm. 
Under this paradigm, Expert Parallelism (EP), a core component of MoE, faces significant scalability issues due to the limited cross-DC bandwidth.
Specifically, existing EP optimizations attempt to overlap data communication and computation, which has little benefit in low-bandwidth scenarios due to a much longer data communication time.
Therefore, the trends of cross-DC EP scaling is fast becoming a critical roadblock to the continued growth of MoE models.

To address this urgent challenge, we propose HybridEP, a modeling-guided framework to optimize EP under constrained bandwidth. 
Our key idea is to dynamically transform the spatial placement of experts to reduce data communication traffic and frequency, thereby minimizing EP's communication overheads.
However, it is non-trivial to find the optimal solution because it complicates the original communication pattern by mixing data and expert communication.
We therefore build a stream-based model to determine the optimal transmission proportion between experts and data. 
Guided by this model, we incorporate two techniques to implement HybridEP: (1) domain-based partition to construct the mapping between hybrid patterns and specific communication topology at GPU level, and (2) parameter-efficient migration to further refine this topology by reducing expert transmission overhead and enlarging the domain size.
Combining all these designs, HybridEP can be considered as a more general EP with better scalability.
Experimental results show that HybridEP outperforms existing state-of-the-art MoE training systems by up to 5.6$\times$ under constrained bandwidth. 
We further compare HybridEP and EP on large-scale simulations.
HybridEP achieves up to 1.45$\times$ speedup with 1000 DCs under different bandwidths.

\end{abstract}

\section{Introduction}
Large language models (LLMs)~\cite{C4, eDKM, BitMoD, Smart-Infinity, TTO} have achieved significant success in various tasks, such as translation~\cite{translation}, text generation~\cite{text-generation}, and question answering~\cite{qa}, driving the community to explore even larger model capacities for better performance. 
Mixture-of-Experts (MoE) has become increasingly popular to enable ultra-scale LLMs. It decouples computation from model size through sparse expert activation, leading to an easy expansion of model parameters to trillions with nearly constant cost~\cite{FSMoE, MoE-Lightning, moe,fastermoe}.

As the scaling law brings about larger pre-training scale, existing training methods in a single data center (DC) face severe challenges~\cite{cross-dc-pp}.
Therefore, recent architectural visions~\cite{cloudmatrix,next-generation-ai-sys} advocate for a more fluid and composable DC infrastructure, where computation, memory, and network resources can be pooled and recombined on demand, forming elastic clusters across smaller, interconnected DCs. 
Driven by this, some pioneers have achieved forward-looking results across DCs~\cite{DiLoCo, opencollaborations} and even countries to scale model size to 10B-32B~\cite{INTELLECT-1,INTELLECT-2}, while maintaining fairly competitive model performance, lower costs, and more flexible resource utilization.
This ongoing paradigm shift makes cross-DC training essential to build the Computing Power Network~\cite{cpn} and make computing power affordable to everyone~\cite{Google, Microsoft, US-utilities}.

Nevertheless, we find that constrained inter-DC bandwidth slows down MoE training, primarily due to Expert Parallelism (EP). EP is the core component of MoE that can account for more than 90\% of training time under low bandwidth, as shown in Figure~\ref{fig:parallelism-analysis}(b). 
Existing EP optimizations~\cite{hetumoe, schemoe, fastermoe, tutel, deepspeedmoe, scomoe, GShard} target a single high-performance DC with the idea of overlapping computation and communication.
% This is not applicable when scaling EP across DC because it is impossible to fully fill all bubbles due to a much longer communication time than computation.
However, they are impractical to scale EP across DCs, as it is impossible to fully overlap computation with the much longer communication time. 
What’s worse, scaling EP across DCs becomes inevitable.
Recent representative MoE models~
\cite{switch-transformer,deepseek,deepseek-v3,mistral-moe,minimax-01,qwen3} have demonstrated a rapidly unfolding trend of EP expansion, bringing an explosive growth of the burden for single-DC training.
Therefore, efficiently scaling EP is a pressing challenge that must be addressed to sustain the future MoE development.

To this end, we propose HybridEP, a modeling-guided framework that can efficiently scale EP under constrained bandwidth.
Our framework tries to answer a more fundamental question: 
\textit{How can we structurally eliminate EP's communication overheads under constrained bandwidth, beyond simply overlapping or hiding them?}
Our key insight is that
\textbf{EP's overheads can be structurally mitigated by adjusting the spatial placement of experts.}
Based on this, HybridEP migrates experts to change their spatial placements, thus altering the communication topology and reducing communication traffic and frequency.
However, such a transformation introduces complex hybrid communication patterns between data and experts, and their proportion has a huge impact on the final effect.
We therefore build a \textbf{Stream-Based Modeling} to decide the best proportion between data and experts.
Our model adopts the divide-and-conquer approach and first it decouples MoE training into two distinct streams (i.e., computation and communication) and independently models them.
Then it analyzes how the two streams overlap and derives a unified end-to-end performance model.

Following this, we implement HybridEP as a practical framework to maximize training efficiency with two techniques:
\textbf{\ding{172} Domain-Based Partition} to construct the hybrid communication topology.
We introduce the expert domain to separate data and expert transmissions, and use a three-step mapping to construct the specific communication topology at the GPU level, ensuring compatibility with existing hierarchical hardware architectures.
\textbf{\ding{173} Parameter-Efficient Migration} to optimize the communication topology.
We further explore redundancy among expert and design a new expert representation to reduce communication traffic, as well as an asynchronous communicator to mitigate synchronization overhead.
With the above designs,  HybridEP
can be considered as a more general EP with better scalability.

Experimental results suggest that HybridEP achieves the minimal training time compared to state-of-the-art MoE training systems. It achieves up to 5.68$\times$ speedup compared to Tutel~\cite{tutel}, FasterMoE~\cite{fastermoe}, and SmartMoE~\cite{scomoe}.
We further conduct a larger scale simulation
to compare HybridEP and EP.
With 1000 DCs connected, HybridEP achieves up to 1.45$\times$
speedup under different bandwidths.

In summary, we make the following contributions:
\begin{itemize}[itemsep=0pt, topsep=0pt, parsep=0pt, leftmargin=10pt]
    \item We find that cross-DC MoE training is the ongoing paradigm to further expand LLM capacity, where EP becomes the main bottleneck to hinder training efficiency.
    
    \item We spatially reshape the placement of experts for efficient EP scaling under constrained bandwidth, which introduces a hybrid communication of data and experts.
    
    \item We implement HybridEP with \textit{Stream-Based Modeling} to get the best proportion of hybrid patterns, \textit{Domain-Based Partition} to construct the specific topology and \textit{Parameter-Efficient Migration} to optimize it for better efficiency. 
    
    \item Experiments suggest that HybridEP outperforms state-of-the-art works in both real scenario and simulation.
\end{itemize}

\section{Background and Motivation}

\begin{figure}[t]
\setlength{\abovecaptionskip}{0.5em}
    \setlength{\belowcaptionskip}{0.5em}
\centerline{\includegraphics{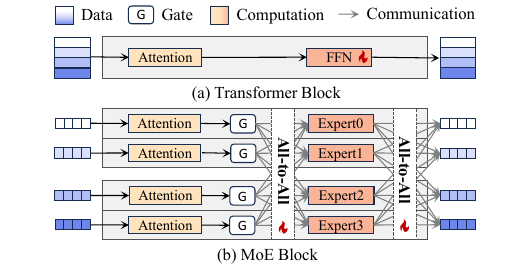}}
\vspace{-0.2cm}
\caption{\textbf{Structures of transformer and MoE block under real training configuration (2 nodes $\times$ 2 GPUs).} 
\emph{Each MoE block replicates FFN to multiple experts, and only activates part of them (through gate network) to process data with constant computation cost. 
However, the frequent communication of All-to-All makes it the main bottleneck during training.}}
\label{MoE-structure}
\vspace{-1em}
\end{figure}

\subsection{Parallelisms, MoE and its Communication Bottleneck}

\textbf{Common Parallelisms.}
There are five common parallelisms during distributed training, each with different properties:

\textit{Tensor Parallelism (TP)}: Splits weights of each layer across multiple GPUs, requiring extensive communication and large traffic during both forward and backward passes.
% Due to limited opportunities for overlap [64] and high communication costs, TP is typically restricted to high-bandwidth domains (e.g., NVLink [40]), .

\textit{Sequence Parallelism (SP)}: Splits sequence across multiple GPUs~\cite{deepspeed-ulysses,ring-attention}, which is typically applied at the end of pre-training for larger context window~\cite{llama3}.

\textit{Pipeline Parallelism (PP)}: Divides the model layers into multiple stages, with each stage assigned to a different GPU. 
It uses point-to-point communication between stages to send/receive of activations and gradients.

\textit{Data Parallelism (DP)}: Replicates the full model on each GPU, where distinct batches are processed independently and
gradients are synchronized across replicas during backward.
% which is also suitable for scaling across DCs.

\textbf{Expert Parallelism (EP)}: Distributes the experts in the MoE model across multiple GPUs.
Specifically, the expert is expanded from the FFN -- a basic module of Transformer structure~\cite{transformer}, as shown in Figure~\ref{MoE-structure}.
On this basis, MoE model further uses a gate network as the router, then uses two \textit{All-to-All} (A2A) communications to change data to the part of activated experts for computation.
However, \uline{large traffic and high frequency of EP makes it almost only deployed in HPC environments currently.}

\begin{figure}[t]
    \centering
    \begin{minipage}[b]{0.48\linewidth}
        % \centering
        \includegraphics[width=\linewidth]{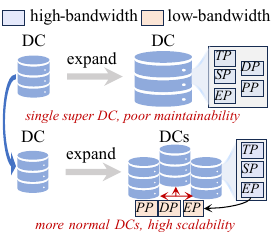}
        \vspace{0.5\baselineskip}
        \small(a) Trend of more DCs,larger EP. \\
        \label{fig:trend}
    \end{minipage}
    \hfill
    \begin{minipage}[b]{0.48\linewidth}
        % \centering
        \includegraphics[width=\linewidth]{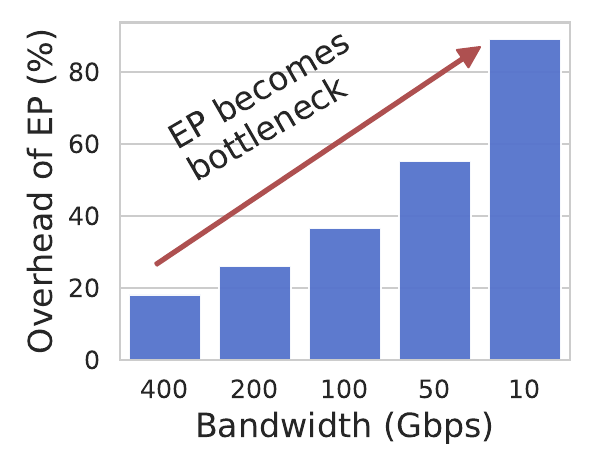}
        \vspace{0.5\baselineskip}
        % \centering
        \small   (b) The overhead ration of EP.  \\
        \label{fig:bottleneck}
    \end{minipage}
    \vspace{-0.7cm}
    \caption{\textbf{Analysis of parallelisms under common settings and the scalability of EP.}
\emph{(a) shows the trends of more DCs and larger EP, which requires more interconnections between DCs and scaling EP across DCs 
(b) shows the overhead ratio of EP under different bandwidths. 
This trend reveals a heavy overhead caused by EP when scaling under low bandwidth.}}
    \label{fig:parallelism-analysis}
    \vspace{-1em}
\end{figure}

\textbf{Necessity of Scaling EP across DCs.}
Firstly, considering the infrastructure, LLMs are pushing existing DCs to their limits. 
However, scaling up a single DC faces challenges like power limitations and increased vulnerability to outages~\cite{Google,Microsoft,US-utilities}. 
Thus, recent reports suggest a more practical and resilient solution is to deploy multiple, smaller-scale DCs~\cite{cross-dc-pp}.
Secondly, considering the model, with the widespread application of MoE, some recent representative MoE models~
\cite{switch-transformer,deepseek,deepseek-v3,mistral-moe,minimax-01,qwen3} have shown a rapidly unfolding trend of expanding number of experts to hundreds or even thousands, exceeding the capacity of a single DC.
Thus, combining the challenges of capacity expansion and the capacity limitations of a single DC, EP will inevitably expand to multiple DCs, as shown in Figure~\ref{fig:parallelism-analysis}(a).

\textbf{EP's Communication Bottleneck.}
% Unfortunately, scaling across DC is extremely inefficient.
As illustrated in Figure~\ref{fig:parallelism-analysis}(b), EP causes extremely serious bottlenecks at relatively low bandwidths.
Existing studies have also shown that under constrained bandwidth, EP almost occupies 50\%-90\% of the overall iteration time~\cite{hetumoe, schemoe, fastermoe}, becoming the main bottleneck in improving training efficiency.
Unlike bandwidth-friendly parallelisms such as DP and PP, the communication-heavy design makes EP particularly sensitive to network conditions.
It relies on the data-to-expert routing pattern (i.e., sending each data to correspond experts via A2A), requiring frequent communication for synchronization across devices and the traffic expands proportionally to the number of activated experts.
Therefore, EP is not designed to scale across DCs and there is an urgent need for efficient EP scaling.
% without sacrificing model performance

\subsection{Motivation and Challenges}
\begin{figure}[t]
\setlength{\abovecaptionskip}{0.5em}
    \setlength{\belowcaptionskip}{0.5em}
\centerline{\includegraphics{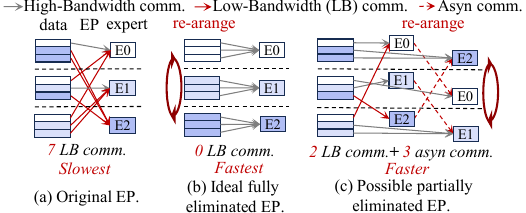}}
\caption{
\textbf{An example of how the spatial placement of data and experts affect EP's efficiency.}
\emph{(a) shows the original placement of EP, where 7 low-bandwidth communications exist with slowest training speed.
(b) shows the ideal case without any low-bandwidth communication through reshaping placements of data, which achieves the best efficiency.
(c) shows a possible optimization in practice with only 4 low-bandwidth communications, achieved only by reshaping the placement of experts.
This achieves a good compromise between ideal case and original EP.} 
}
\label{opportunity}
\vspace{-1em}
\end{figure}

In Figure~\ref{opportunity}, we use a simple example to show how the placement of data and experts affects EP's communication overheads and motivates HybridEP.

\textbf{Case of Original EP.}
In Figure~\ref{opportunity}(a), the basic logic of original EP operates like sending data to the corresponding experts decided by the routing result.
When experts are placed across different nodes connected by low bandwidth, this introduces multiple slow communications (e.g., 7 in our example), which becomes the main bottleneck and degrades EP's efficiency.

\textbf{Case of Ideal Fully-Eliminated EP.}
% In Figure~\ref{opportunity}(b), 
Ideally, if we know in advance which expert the data will be routed to, we can re-arrange before training and place data sent to the same expert together with high-bandwidth interconnections, avoiding the huge low-bandwidth communication latency, as shown in Figure~\ref{opportunity}(b).
This essentially moves EP to the beginning of training, reducing its frequency and traffic and achieving the fastest speed.
However, the problem is that the data routing results are dynamic and determined in real time, so it is almost impossible to know the routing results in advance and re-arrange the data before training. 
Meanwhile, since data-to-expert assignments vary across layers, it is unrealistic to assume the perfect placement before training.
Thus, it is almost impossible to achieve this.
% Meanwhile, since data at different layers often have different routing results, it is difficult to unify them before training.
% However, imagine an ideal mapping where each GPU processes all data that consistently route to the same expert. 
% If we could relocate that expert's weight onto the GPU hosting those tokens beforehand, then token activations no longer need cross-device transmission — they are processed locally, completely eliminating A2A overhead for those tokens.

\begin{figure}[t]
\setlength{\abovecaptionskip}{0.5em}
    \setlength{\belowcaptionskip}{0.5em}
\centerline{\includegraphics[width=\linewidth]{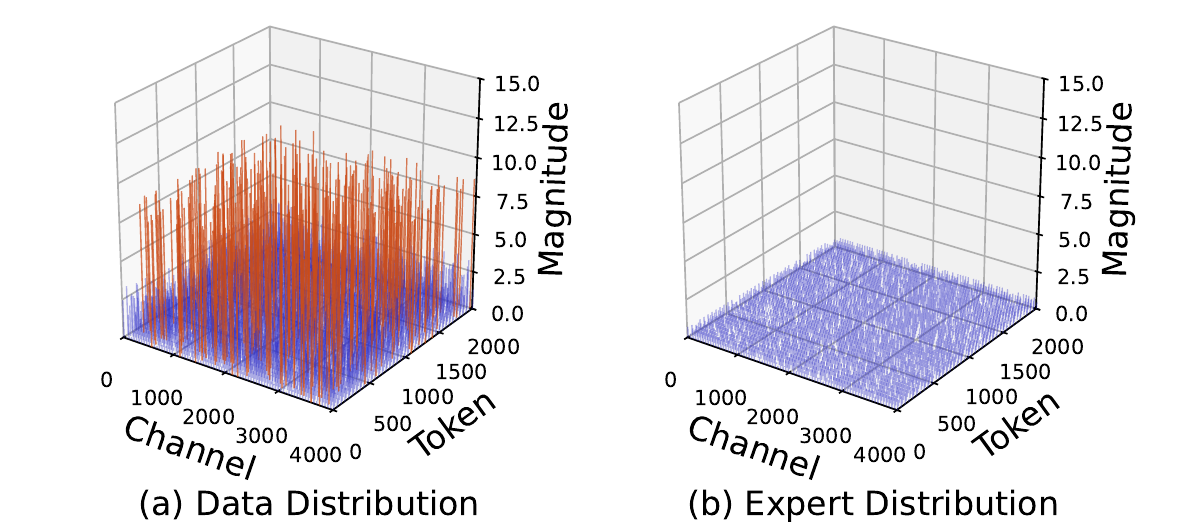}}
\caption{
\textbf{Compressibility analysis of data and experts.}
\emph{The distribution of data has large magnitude and many outliers (red part), while the distribution of expert weight is relative small and flat, leading to a higher compressibility.}
% \emph{(a) shows that expert has high compressibility because its distribution (E) is more centralized than data's (D1-D5). Subfigure (b) shows that residual experts are more centralized than original experts with even higher compressibility.}
}
% \caption{The distribution of data and expert (a) and the redundancy among all experts (b).}
\label{Compressibility}
\vspace{-1em}
\end{figure}

\textbf{Case of Possible Patially-Eliminated EP.}
Considering that the basic logic of EP is to assign data with corresponding experts, we can optimize EP efficiency by changing the spatial placement of experts, as shown in the Figure~\ref{opportunity}(c).
After experts process their data, we re-arange the placement of experts and then continue the computation.
Although it can not achieve the ideal case, it still reduces the number of low-bandwidth communications (e.g., 5 in our example) for better efficiency.
Note that we distinguish the transmission of data and expert, because we find that the expert has two rather good advantages for lightweight transmission compared to data, which has not been fully exploited by previous works to optimize the efficiency of EP.
Specifically, \textit{\ding{172} Better compressibility.}
Expert weights typically exhibit compact and concentrated distributions, with fewer outliers compared to data, as shown in Figure~\ref{Compressibility}, which is also confirmed by prior works~\cite{smoothquant, AWQ, KIVI, KVQuant}.
This allows experts to be compressed more aggressively without degrading accuracy to reduce traffic.
\textit{\ding{173} Asynchronous communication potential.} 
Unlike data that are tightly coupled with every computational operator, experts only participate in computation intermittently.
This allows to pre-transmit experts independently ahead of EP, reducing synchronization overhead.
Together, these points underscore that experts are inherently suited for lightweight, early, and stable communication.
This forms the foundation for HybridEP.

\textbf{Why this works?}
The key reason is that experts are \textit{sparsely activated}, which is also the biggest difference between EP and the remaining parallelisms.
Only in EP, data are processed by a small number of experts (part of model parameters), while in other parallel modes, all data need to be processed by all model parameters. 
Analogously to Figure~\ref{opportunity}(c), 
if EP does not have the property of sparse activation, that is assuming each data needs to be calculated with all experts, then no matter how the expert placement is changed, the number of low-bandwidth EP communications will not change.
Therefore, only parallelisms with sparse activation properties like EP can improve efficiency through changing expert placement.
% Therefore, we make full use of this feature to efficiently scale EP.

\textbf{Challenges.}
The above analysis provides a vision of efficient EP scaling, introducing a hybrid transmission of both experts and data. 
However, in a real training scenario, it is still non-trivial to maximize its potential, which can be summarized into three main challenges:
\begin{itemize}
    \item \textit{\ding{172} How to determine the best proportion of data-expert hybrid communication patterns?}
    Essentially, experts communicate in All-Gather (AG) pattern, as each device collects experts from others (details in \S\ref{modeling}). 
    We need to balance this with the A2A pattern for data.
    
    \item \textit{\ding{173} How to construct the specific communication topology and ensure compatibility with existing architecture?} 
    Even if the proportion of two communications is determined, the most basic communication granularity is GPU, not DC.
    Thus a specific topology needs to be built to align the communication granularity.

    \item \textit{\ding{174} How to effectively exploit high compressiblity and asynchronicity for lightweight expert migration?}
    Although expert transmission can be lightweight, we still need to design corresponding compression and communication mechanisms to achieve theoretical efficiency.
\end{itemize}
To this end, we propose HybridEP to address the above challenges.
First, we design a stream-based modeling to address the first challenge and provide a high-level guidance for the overall design. 
Following this, we introduce two techniques to unlock its potential. 
We address the second challenge by determining the hybrid communication topology and compatibility with existing hierarchical architecture through the division of expert domains.
Then we address the third challenge by designing an expert compression algorithm and an asynchronous communicator for lightweight expert migration.

\section{Foundation: Stream-Based Modeling}
\label{modeling}

\begin{table}[t]
\setlength{\abovecaptionskip}{0.5em}
    \setlength{\belowcaptionskip}{0.5em}
\caption{\textbf{Frequently Used Notations in Modeling.}}
\begin{center}
\small
\begin{tabular}{c|l}
\hline
\textbf{Name}& \textbf{Description}\\
\hline\hline
$D$ & Data size of a GPU. \\
$P_E$ & Expert size with \# parameters. \\
$C$ & Computation throughput of a GPU. \\
$B$ & Communication bandwidth between GPUs.\\
$V^x$ & Communication volume of $x$. $x$ is either A2A or AG. \\
$Lat_{x}^{y}$ & $y$'s latency of operation $x$ (either comp. or comm.).\\
$G$ & Number of GPUs in the cluster. \\
$G^x$ & GPU set $x$. $x$ is either A2A or AG.\\
$|G^x|$ & Number of GPUs in GPU set $G^x$. \\
\hline
\end{tabular}
\label{tab:notation}
\vspace{-1em}
\end{center}
\end{table}

The modeling process is shown in Figure~\ref{comm_breakdown}(a).
We first split MoE training into computation and communication streams (\S\ref{subsec:computation-modeling}, \S\ref{subsec:communication-modeling}).
Then, we model the overlap between two streams (\S\ref{subsec:overlapping-modeling}) and jointly consider computation, communication, and overlap to minimize training latency (\S\ref{subsec:problem-formulation}, \S\ref{subsec:problem-solving}).
We summarize frequently used notations in Table~\ref{tab:notation}.
Our modeling assumes that each GPU has the same expert number, each expert has the same size, and the gate network activates experts evenly.
For simplicity, we first assume that there is only one GPU in each DC (using GPU to represent the DC) to align the communication granularity, which does not affect the accuracy of our model.

\subsection{Computation Modeling}\label{subsec:computation-modeling}

\textbf{GeMM Modeling.}
The computation latency mainly comes from General Matrix Multiplication (GeMM) operations.
Following prior works~\cite{fastermoe, GPU-time-estimate}, we use a linear model to estimate the latency.
Given two matrices to be multiplied, with size $(L, H)$ and $(H, M)$ respectively, 
% Given two a data matrix of size $(L, H)$ and a parameter matrix of size $(H, M)$, GeMM requires $BHM$ Fused Multiply-Add (FMA) operations.
% Therefore, given a GPU with average computation throughput $C$, 
the latency of a single GeMM operation can be expressed as:
\begin{equation}
    Lat_{comp}^{GeMM} = \frac{LMH}{C},
    \label{eq:gemm-lat}
\end{equation}
where $C$ represents the average computation throughput of GPU.
Note $C$ will be reduced if GeMM is too small to utilize GPU power. 
However, this will not affect overall modeling effectiveness due to its small overhead, confirmed by~\cite{fastermoe}.

\textbf{Computation Stream Modeling.}
% MoE models consist of transformer blocks and MoE blocks. 
% For ease of presentation, we 
Assume that there are $m$ transformer blocks before a MoE block, 
% which can be changed for different configuration. 
The computation latency can be expressed as
\begin{alignat}{2}
    Lat_{comp} & = mLat_{comp}^{TF} + Lat_{comp}^{MoE} \nonumber \\
    & = (m+1)Lat_{comp}^{Att} + mLat_{comp}^{FFN}+nLat_{comp}^{Ep},
    \label{eq:comp-lat}
\end{alignat}
where $TF, MoE$ represents the transformer and MoE block.
$Att, FFN, Ep$ represents the computation process of attention, FFN, and expert, which consists of multiple GeMM operations. Here we can consider their latency as constant.
$n$ is the number of experts on one GPU, thus expert computation latency is repeated by $n$ times.
For brevity, we consider $(m+1)Lat_{comp}^{Att} + mLat_{comp}^{FFN}$ as \textit{Pre-Expert}, denoted $Lat_{comp}^{PE}$.

\subsection{Communication Modeling}\label{subsec:communication-modeling}

\begin{figure}[t]
\setlength{\abovecaptionskip}{0.5em}
    \setlength{\belowcaptionskip}{0.5em}
\centerline{\includegraphics{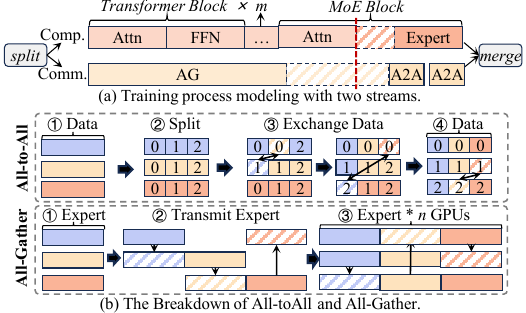}}
\vspace{-0.2cm}
\caption{
\textbf{The modeling of training process 
and the communication breakdown of A2A and AG.} \emph{(a) A shows a modeling process using the divide-and-conquer approach. The training process is first split into independent modeling of computing and communication streams, and then their overlapping relationships are considered for merging.
(b) shows that the traffic of A2A remains unchanged (i.e., O(1)), while the traffic of AG is multiplied by number of GPUs (i.e., O(n)).}}
% The breakdown of all-to-all and all-gather communication for traffic analysis ($G=3$).
\label{comm_breakdown}
\vspace{-1em}
\end{figure}

\textbf{All-to-All Communication Modeling.}
Figure~\ref{comm_breakdown}(a) shows A2A communication details~\cite{split-a2a}.
Specifically, given $G$ GPUs, the data $D$ on each GPU will be split into $G$ chunks of size $\frac{D}{G}$. Then, $G-1$ chunks will be sent to other GPUs through A2A, while $1$ chunk remains on the local GPU.
Therefore, for GPU set $G^{A2A}$ that participates in A2A communication, the overall traffic is expressed as:
\begin{alignat}{2}
    V^{A2A} = \frac{D}{|G^{A2A}|} * (|G^{A2A}|-1), Lat_{comm}^{A2A}=\frac{V^{A2A}}{B}, \label{eq:a2a-vol}
\end{alignat}
where $B$ is bandwidth.
While $D,B$ are constants, \uline{$Lat_{comm}^{A2A}$ remains almost constant with increased $|G^{A2A}|$.}

\textbf{All-Gather Communication Modeling.}
Figure~\ref{comm_breakdown}(b) shows AG communication details.
Specifically, the expert parameters $P_E$ on each GPU will be sent to other $G-1$ GPUs through AG.
Therefore, for GPU set $G^{AG}$ that participates in AG, communication traffic is expressed as:
\begin{alignat}{2}
    V^{AG} &= P_E * (|G^{AG}|-1), Lat_{comm}^{AG} &= \frac{V^{AG}}{B} .\label{eq:ag-vol}
\end{alignat}
% Given bandwidth $B$, AG latency can be expressed as:
% \begin{alignat}{2}
%     Lat_{comm}^{AG} &= \frac{V^{AG}}{B} = P_E * \frac{|G^{AG}|-1}{B}. \label{eq:ag-lat}
% \end{alignat}
\uline{Therefore, $Lat_{comm}^{AG}$ increases linearly with $|G^{AG}|$.}

\textbf{Relationships between Two Communications.}
A2A can be seamlessly transformed into AG. 
If an expert has been obtained through AG, then the corresponding data chunk is not necessary to be transmitted through A2A.
% For example, given the $i$-th GPU $G_i$, the vanilla data traffic is $V^{A2A}=D*\frac{G-1}{G}$ and $G^{A2A}$ contains all GPUs, while $G^{AG}$ only contains $G_i$ with traffic $V^{AG}=0$.
When the $i$-th GPU $G_i$ uses AG to collect expert $P_E$ from $G_j$,
the A2A's traffic changes from $D*\frac{G-1}{G}$ to $D*\frac{G-1}{G}-\frac{D}{G}$, while the AG's traffic changes from $0$ to $P_E$.
% this brings about two changes:
% 1. The A2A GPU set changes from $G^{A2A}$ to $G^{A2A}-\{G_j\}$, and the data traffic changes from $D*\frac{G-1}{G}$ to $D*\frac{G-1}{G}-\frac{D}{G}$.
% 2. The AG GPU set changes from $G^{AG}$ to $G^{AG}+\{G_j\}$, and the expert traffic changes from $0$ to $P_E$.
Therefore, \uline{when A2A's traffic decreases by $\frac{D}{G}$, AG's traffic increases by $P_E$.}

\textbf{Communication Stream Modeling.}
Its latency comes from both A2A and AG, which can be expressed as:
\begin{alignat}{2}
    Lat_{comm} & = Lat_{comm}^{AG} + 2Lat_{comm}^{A2A}. 
    % \nonumber \\
    % & = \frac{V^{AG}}{B} + 2 \frac{V^{A2A}}{B}.
    \label{eq:comm-lat}
\end{alignat}
where A2A performs twice before and after expert computation, and AG only performs once as experts do not need to be sent back to their original GPUs.

\subsection{Overlap Modeling}\label{subsec:overlapping-modeling}

\textbf{Overlap Modeling of Two Streams.} 
The overlap time between computation and communication comes from three parts, as shown in Figure~\ref{comm_breakdown}(a) split by the red dotted line.
Specifically, \ding{172} pre-expert computation (i.e., $Lat_{comp}^{PE}$) and AG; \ding{173} expert computation ($Lat_{comp}^{Ep}$) and AG; \ding{174} expert computation ($Lat_{comp}^{Ep}$) and A2A; 
Note that the pre-expert computation cannot overlap with A2A because the data depend on the pre-expert results.
Therefore, overlap can be written as:
\begin{equation}
    Lat_{ovlp} = Lat_{ovlp}^{PE,AG} + Lat_{ovlp}^{Ep,AG} + Lat_{ovlp}^{Ep,A2A},
\end{equation}
Note that case \ding{173} and \ding{174} have been optimized by previous works~\cite{pipemoe, Janusmoe}, therefore expert computation is fully overlap with AG and A2A ($Lat_{ovlp}^{Ep,AG} + Lat_{ovlp}^{Ep,A2A} = nLat_{comp}^{Ep}$).
For case \ding{172}, the overlap time is $min(Lat_{comp}^{PE},Lat_{comm}^{AG})$.
Therefore, the final overlap time can be expressed as: 
\begin{equation}
    Lat_{ovlp} = min(Lat_{comp}^{PE},Lat_{comm}^{AG}) + nLat_{comp}^{EP},
    \label{eq:overlap-lat}
\end{equation}

\subsection{Problem Formulation}\label{subsec:problem-formulation}

% The overall training latency is the sum of the computation and communication latency, minus the overlap latency. 
% To minimize the latency, we define the proportion of A2A and AG :
To minimize the latency, we have the following definition:

\textbf{Definition 1.} 
\textit{Given a cluster with $G$ GPUs ($G>1$), $p_i$ is the proportion of data chunks (which leave from $G_i$) that are transmitted through A2A, while $1 - p_i$ is the proportion of data chunks (which leave from $G_i$) that are transformed into expert and transmitted through AG, where $p_i \in\{\frac{0}{G-1},\cdots,\frac{G-1}{G-1}\}$.}

When $p_i=\frac{G-1}{G-1}$, there is only A2A; when $p_i=\frac{0}{G-1}$, there is only AG. The training latency can be expressed as: 
\begin{alignat}{2}
    \min_{p_i} \quad & Lat_{final}(p_i) = Lat_{comp} + Lat_{comm} - Lat_{ovlp}     \label{eq:overall-lat} \\
    \mathrm{s.t.} \quad & p_i\in \{\frac{0}{G-1},\cdots,\frac{G-1}{G-1}\}, Eq~\ref{eq:comp-lat},Eq~\ref{eq:comm-lat} ,Eq~\ref{eq:overlap-lat}. \nonumber
\end{alignat}
% ,Eq~\ref{eq:gpu-constrain2} ,Eq~\ref{eq:gpu-constrain}

Note that each GPU has its own Eq~\ref{eq:overall-lat}, and they should be synchronized. Therefore, system latency is the maximal latency of all GPUs, which can be expressed as:
\begin{equation}
    Lat_{all}=\max_{0\leq i<G} \{\min Lat_{final}(p_i)\}
    \label{eq:lat-all}
\end{equation}
\uline{Finally, Eq~\ref{eq:lat-all} depends solely on parameter $p_i$, and our goal is to minimize $Lat_{all}$ by choosing the optimal $p_i$.}

\subsection{Problem Solution}\label{subsec:problem-solving}
For simplicity, we assume that all the $p_i$ are the same.
Thus, Eq~\ref{eq:lat-all} can be simplified to an easy-to-solve format:
% GPUs and communication bandwidth are homogeneous with the same latency. Therefore, each GPU has the same training latency when they have the same $p_i$, and Eq~\ref{eq:lat-all} can be simplified to:
\begin{align}
    &Lat_{all} = \min Lat_{final}(p) \nonumber \\
    % =&\min (Lat_{comp}^{PE} + Lat_{comm} - min(Lat_{comp}^{PE},Lat_{comm}^{AG})) \nonumber\\
    =&
    \begin{cases} 
        \min (Lat_{comp}^{PE}+2Lat_{comm}^{A2A}),\text{if } Lat_{comp}^{PE} \geq Lat_{comm}^{AG} \\
        \min (Lat_{comm}^{AG}+2Lat_{comm}^{A2A}),\text{if } Lat_{comp}^{PE} < Lat_{comm}^{AG}. \\
    \end{cases}
    \label{eq:overall-lat-simple}
\end{align}
The final solution can be organized into two cases.

\noindent \textbf{Case 1:}when $Lat_{comp}^{PE} \geq Lat_{comm}^{AG}$, Eq~\ref{eq:overall-lat-simple} is simplified as:
\begin{alignat}{2}
    \begin{cases}
        Lat_{all} &= Lat_{comp}^{PE} + \frac{2D(G-1)}{GB}p  \\
        \frac{G-1}{G-1} &\geq p \geq \frac{P_E(G-1)-BLat_{comp}^{PE}}{P_E(G-1)}
    \end{cases}
    \label{eq:case1-overall-lat}
\end{alignat}
Note that $Lat_{comp}^{PE}, D, B$ are positive constants. 
% therefore $Lat_{final}$ increases as $p$ increases. 
Thus, to minimize $Lat_{final}$, we need to configure the minimum $p$.

\begin{figure}[t]
    \setlength{\abovecaptionskip}{0.5em}
    \setlength{\belowcaptionskip}{0.5em}
    \centering
    \includegraphics[width=0.9\linewidth]{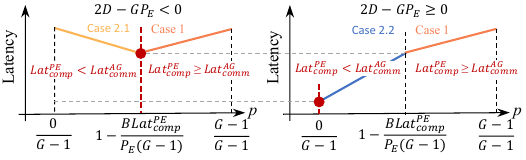}
    % \caption{The trend of the overall latency $Lat_{final}$ as a function of $p$.}
    \vspace{-0.2cm}
    \caption{\textbf{Visualization of Eq~\ref{eq:overall-lat-simple}'s solution.} \emph{Two red dots indicate the optimal $p$ with minimal latency under two cases.}}
    \label{function}
    \vspace{-1em}
\end{figure}

\noindent \textbf{Case 2:}when $Lat_{comp}^{PE} < Lat_{comm}^{AG}$, Eq~\ref{eq:overall-lat-simple} is simplified as
\begin{alignat}{2}
    \begin{cases}
        Lat_{all} &= p\frac{(G-1)(2D-GP_E)}{BG}+\frac{(G-1)P_E}{B}  \\
        \frac{0}{G-1} &\leq p < \frac{P_E(G-1)-BLat_{comp}^{PE}}{P_E(G-1)}
    \end{cases}
    \label{eq:case2-overall-lat}
\end{alignat}
Note that the sign of $\frac{(G-1)(2D-GP_E)}{BG}$ has two cases for minimal $Lat_{final}$.
% cannot be directly determined to positive or negative (i.e., the sign of coefficient of $p$), where we need further discussions:
\ding{172} When $2D-GP_E < 0$, 
% $Lat_{final}$ increases as $p$ decreases, therefore 
we configure the maximum $p$, denoted \textbf{Case 2.1}.
\ding{173} When $2D-GP_E \geq 0$, 
% $Lat_{final}$ increases as $p$ increases, therefore 
we configure the minimum $p$, denoted \textbf{Case 2.2}.

\textbf{Summary.}
Our model find the best proportion for minimal latency (i.e., $p$), which can be summarized into two cases, as shown in Figure~\ref{function}.
Specifically, when $2D-GP_E<0$, the variation of overall latency consists of Case 1 and Case 2.1.
Therefore, the optimal $p$ is configured to $1-\frac{BLat_{comp}^{PE}}{P_E(G-1)}$, where we use both AG and A2A.
When $2D-GP_E\geq0$, the variation of overall latency consists of Case 1 and Case 2.2. 
Therefore, the optimal $p$ is configured to 0, where we only use AG.
Note that when $p=1$, HybridEP degenerates into the standard EP, indicating that EP is a special case of our framework.
\section{Design and Implementation}

\begin{figure}[t]
    \setlength{\abovecaptionskip}{0.5em}
    \setlength{\belowcaptionskip}{0.5em}
    \centering
    \includegraphics{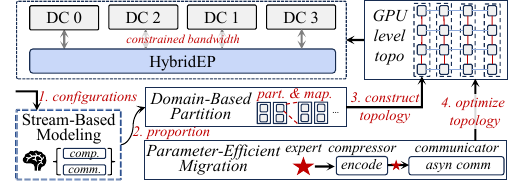}
    \vspace{-0.5cm}
    \caption{\textbf{HybridEP overview.} 
    \emph{After modeling decides the proportion of transmitting data and expert, HybridEP uses the domain-based partition to construct specific GPU communication topology. 
    Moreover, the parameter-efficient migration reduces the overhead for a better partition.}}
    \label{overview}
    \vspace{-1em}
\end{figure}

The overview of HybridEP is shown in Figure~\ref{overview}.
Before training, HybridEP first takes the environmental configurations as input and uses the modeling to find the best proportion of transmitting data and experts. 
Oriented by this, HybridEP then introduces \emph{domain-based partition} to partition GPUs for A2A and AG communication (\S\ref{subsec:domain-management}), which constructs the communication topology at GPU level.
% When the expert domain is determined, how communication (AG and A2A) overlaps with computation (pre-expert, including attention and FFN layers and expert) is also settled. 
Moreover, HybridEP designs parameter-efficient migration to optimize the determined communication topology with a better partition (\S\ref{efficient-comm}).

\subsection{Domain-Based Partition}\label{subsec:domain-management}

\begin{figure}[t]
    \setlength{\abovecaptionskip}{0.5em}
    \setlength{\belowcaptionskip}{0.5em}
    \centering
    \includegraphics{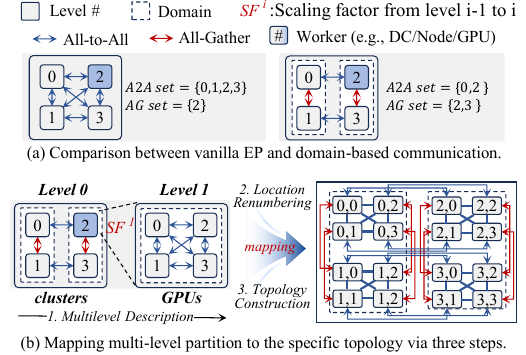}
    \vspace{-0.5cm}
    \caption{\textbf{Domain-based communication and the topology construction at multilevel.} \emph{(a) shows how expert domain affects communication, which splits communications into the in-domain AG and the cross-domain A2A.
    (b) shows the mapping between topology and multilevel partition through three key steps: \textit{Multilevel Description}, \textit{Location Renumbering} and \textit{Topology Construction}.
    % (a) shows the vanilla MoE only uses A2A for communication, we leverage domain to use both A2A and AG. 
    % Moreover, it also achieves hierarchical communication to adapt to heterogeneous bandwidth.
    }}
    \label{group-split}
    \vspace{-1em}
\end{figure}

\textbf{Expert Domain and the Domain-Based Communication.}
The \textit{Expert Domain} is a set of DCs that only uses AG communication within it.
% denoted $G^{AG}$, 
The size of expert domain is defined as the number of DCs in it, denoted as $S_{ED}$. 
HybridEP assume that each domain has the same size.
Figure~\ref{group-split}(a) right side shows an example, we set $S_{ED}=2$ and sequentially group every 2 DCs into the same domain. 
With the help of expert domain, we have the following domain-based communication rule: \uline{AG will only occur for intra-domain communication, and A2A will only occur for inter-domain communication.}
Such a simple rule can effectively separate the two communication patterns for better management.

% \textbf{Scaling Domain to Multi-Level for the Hierarchical Architecture.}
\textbf{Necessity of Scaling to Multilevel.}
In actual scenario, the training environment often consists of hierarchical architectures, and the basic communication granularity is GPU. 
Thus, although how to communicate between DCs is determined, the specific behavior of each GPU is still unclear.
Aligning with the GPU granularity is a critical step for real training scenarios.
To bridge this gap, HybridEP first abstracts the hierarchical structure into \textit{Multilevel Description}, handling the complex and changeable environments in reality.
Then, it renumbers the global GPU number via \textit{Location Renumbering} to adapt to the multilevel.
Finally, it performs the \textit{Topology Construction} algorithm to determine the specific topology at GPU level.
% the final communication topology is constructed by GPU  and \textit{}.
The workflow is illustrated in Figure~\ref{group-split}(b). 

\textbf{Multilevel Description.}
We first define that \emph{Worker} is a physical entity (e.g., DC, node, or GPU).
Normally, we consider GPU as the smallest granularity of a worker.
\emph{Level} is a set of workers that are connected with homogeneous bandwidth. 
Thus, we expand the definition of expert domain size at level $l$ as the number of workers in the domain, denoted as $S_{ED}^l$. 
% All levels have their own expert domains, and we use expert domain size $S_{ED}^l$ to record the number of workers in the domain of level $l$.
To describe the relationship between different levels, we use the \emph{scaling factor} $SF^i$ to indicate that a worker at level $i-1$ can be expanded to level $i$ with $SF^i$ sub-workers.
% Therefore, given the scaling factor $SF^i$, level $i$ has $SF^i$ workers in total.
Note that we set $SF^0$ to the total number of workers at level 0.
Take Figure~\ref{group-split}(b) as an example, given an environment with 4 DCs and each with 4 GPUs, it is split into two levels with $SF^0=4, SF^1=4$ and the domain size at each level is $S_{ED}^0=2,S_{ED}^1=4$, respectively.
% , and the bandwidth within a level is homogeneous.

\textbf{Location Renumbering.}
To clarify detailed communication rules, we first renumber the locations for each GPU for multi-level architecture. 
Specifically, we follow Pytorch\cite{pytorch} to allocate a global index $m$ to each GPU. 
Then, given a $L-1$ level partition, we renumber the global index $m$ into multi-level locations $(x_0,x_1,\cdots,x_{L-1})$.
With the scaling factor list $[SF^0,\cdots,SF^{L-1}]$, the renumbering function $f: m \mapsto (x_0,x_1,\cdots,x_{L-1})$ can be expressed as:
\begin{alignat}{2}
    f(m)= & (x_0,x_1,\cdots,x_{L-1}) \nonumber \\
    x_i = \frac{m}{\prod_{j=i+1}^{L-1} SF^j} & \mod SF^i , i\in\{0, 1, \cdots,L-2\} \nonumber\\
    x_{L-1}=m &\mod SF^{L-1}
\end{alignat}
Therefore, GPU $m$'s level-$i$ worker number can be obtained by $f(m)[i]$. Moreover, with the expert domain size $S_{ED}^i$, GPU $m$'s level-$i$ domain can be obtained by $\frac{f(m)[i]}{S_{ED}^i}$.

\textbf{Topology Construction.}
The related pseudocode is shown in Algorithm~\ref{alg:multi-level-route}. 
Specifically, given two GPUs with global index $m$ and $n$, we decide which type of communication is required at different levels.
We first obtain their multilevel locations by $f(m)$ and $f(n)$. 
To limit the inefficiency caused by multiple communications, we 
% formulate communication rules to
limit the range of GPUs that can communicate with each other.
Specifically, only when $f(m)[l]\neq f(n)[l]$ and the indices of subsequent layers are the same, two GPUs can communicate with each other.
At each level, the communications between GPUs follow the domain-based communication rule.

\begin{algorithm}[t]
\caption{\textbf{Communication Topology Construction}} \label{alg:multi-level-route}
% Communication Path of Expert and Data
\begin{algorithmic}[1]
\STATE \textbf{Input:} GPU $m$, GPU $n$, current level $l$, scaling factor $SF^l$, expert domain size $S_{ED}^l$
\STATE \textbf{Output:} Communication type (None or AG or A2A)
\STATE $Loc_m \leftarrow f(m)$
\STATE $Loc_n \leftarrow f(n)$
\STATE $W_m,W_n \leftarrow Loc_m[l],Loc_n[l]$ 
\STATE $ED_m,off_m \leftarrow \frac{W_m}{S_{ED}^l},W_m \mod S_{ED}^l$
\STATE $ED_n,off_n \leftarrow \frac{W_n}{S_{ED}^l},W_n \mod S_{ED}^l$
\IF{$Loc_D[l+1:]==Loc_E[l+1:]$}
    \IF{$ED_n==ED_m$ and $off_n \neq off_m$}
        \RETURN AG
    \ENDIF

    \IF{$ED_D \neq ED_E$ and $off_D == off_E$}
        \RETURN A2A
    \ENDIF
\ENDIF
\RETURN None
\end{algorithmic}
\end{algorithm}

\begin{figure}[t]
    \centering
    % 第一个子图
    \begin{minipage}[b]{\linewidth} % 第一个子图区域，宽度可以更宽
        \centering
        \includegraphics[width=\linewidth]{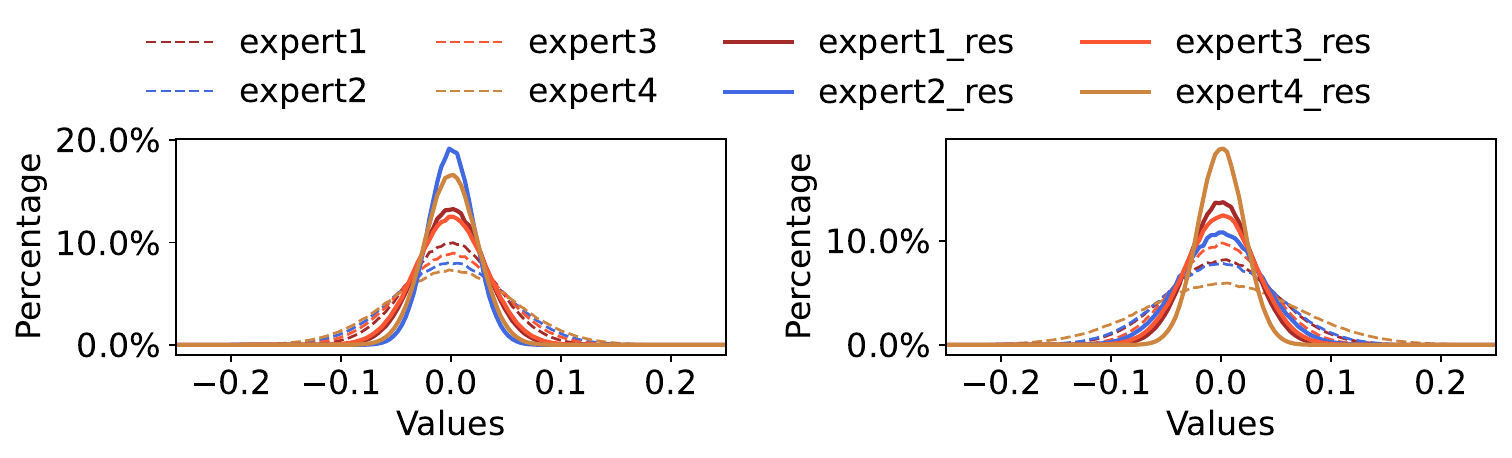}
        \vspace{0.5\baselineskip}
        \small (a) Both two weight matrices of experts has redundancy. \\
        \label{fig:redundancy}
    \end{minipage}
    \par % 强制换行
    % \vspace{4cm} % 可选：增加垂直间距

    % 第二个子图
    \begin{minipage}[b]{\linewidth} % 第二个子图区域
        \centering
        \includegraphics{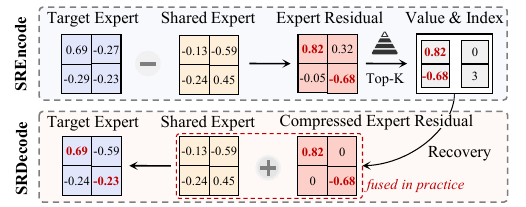}
        \vspace{0.5\baselineskip}
        \small (b) Two phases of compressing and decompressing experts.  \\
        \label{fig:subim2}
    \end{minipage}
    \vspace{-0.8cm}
    \caption{\textbf{Redundancy amoung experts and the workflow of SR-Based Expert Compression.} \emph{
    % S means Shared Expert, while R means Expert Residual.
    In SRDecode, we fuse the recovery and the addition operation in practice for better efficiency.}}
    \label{SRexpert}
    \vspace{-1em}
\end{figure}

\subsection{Parameter-Efficient Migration}
\label{efficient-comm}

% \textbf{How Compressor Works in HybridEP.}
\textbf{How Lightweight Migration Optimizes Communication Topology.}
Essentially, the lightweight migration reduces the size of $P_E$, leading to a larger expert domain which achieves better efficiency.
Specifically, as shown in Figure~\ref{SRexpert}, a smaller $P_E$ can lead to a smaller $p$ mainly due to two aspects: 
1. When $2D<GP_E$, the corresponding $p$ of the optimal point (i.e., the red dot) will decrease.
2. It allows more training configurations to be converted from $2D<GP_E$ to $2D\geq GP_E$.
A smaller $p$ indicates a larger domain of experts, which changes the constructed communication topology.
Furthermore, the larger domain, the better efficiency can achieve.
This is because Eq~\ref{eq:case1-overall-lat} and Eq~\ref{eq:case2-overall-lat} show that the overall latency decreases after $p$ decreases theoretically.
Thus, we regard parameter-efficient migration as a process of optimizing the communication topology by expanding the domain of experts, which aims to further improve the efficiency.

\textbf{Redundancy Among Experts}
In addition to the better compressibility of experts shown in Figure~\ref{Compressibility}, we further explored the redundancy among experts to improve compression ratio.
We find that the main differences among experts are concentrated in a small number of parameters. 
It suggests that different experts may learn similar knowledge from data, which is also reported in other related work~\cite{deepseek}. 
% We trained the GPT-S model for one epoch on a cluster with 4 GPUs and recorded expert weights on each GPU. 
As shown in Figure~\ref{SRexpert}(a), after averaging expert weights and subtracting them from the original weights, the result's distribution (with suffix "res") is more concentrated than the originals.
This indicates that the residuals are sparse and the key differences between experts focused on a few parameters.

\textbf{SR-Based Expert Compression.} 
% Observing that expert residuals are more centralized than experts (Figure~\ref{Compressibility}(b)), 
We are motivated to divide experts into shared and residual parts, which learn redundant knowledge and specific knowledge separately. 
Specifically, the shared expert is shared by all GPUs and is initialized by averaging all experts. At each training iteration, it will be synchronized with asynchronous All-Reduce in the backward propagation phase.
Our expert compression has two phases, as shown in Figure~\ref{SRexpert}(b).
In the \textit{encode phase}, the compressor first obtains the expert residual by subtracting the target expert and the shared expert. Then, it compresses expert residual through Top-k. The compressed expert residual is saved in the value-index format to transmit to other GPUs. 
% Note that encode will be launched immediately when updating optimizer at the beginning of each iteration.
In the \textit{decode phase}, the compressor first recovers the compressed expert residual. Then, it restores the target expert by adding up the shared expert and the residual expert.
Note that in practice, we fused the above two steps of decode phase for less overhead.

\begin{figure}[t]
    \setlength{\abovecaptionskip}{0.5em}
    \setlength{\belowcaptionskip}{0.5em}
    \centering
    \includegraphics{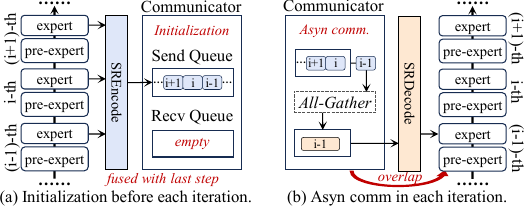}
    \vspace{-0.5cm}
    \caption{\textbf{Tow stage of asynchronous communicator.} \emph{(a) shows the initialization stage, which is fused with the last optimizer step. 
    Each MoE layer sequentially sends their experts processed by SREncode to Send Queue.
    (b) shows the asyn comm stage, which is overlapped with pre-expert computation.
    The communication results of
    each MoE layer are stored in Recv Queue and processed by SRDecode for subsequent computation.}}
    % \caption{Two phases of SR expert representation. In SRDecode, we fuse the decompression and the addition operation in practice to save memory.}
    \label{fig:asynchronous}
    \vspace{-1em}
\end{figure}

\textbf{The Mechanism of Asynchronous Communicator.} 
% Considering the flexibility of experts, 
We use an asynchronous communicator to achieve the theoretical effect of our modeling as much as possible.
To fully combine the asynchronous characteristics with SR compression without too much extra overhead, 
% To minimize the impact on the theoretical improvement in our modeling as much as possible, 
we divide the behavior of the asynchronous communicator into two stages like SR compression.
As illustrated in Figure~\ref{fig:asynchronous}, the communicator considers the model as a stack of (pre-expert, expert) pair, and has a Send Queue and a Recv Queue.
In \textit{Initialization} stage, all experts in the model are sequentially processed by SREncode, and the compressed results will be delivered to the Send Queue.
The Recv Queue is set to be empty.
Note that this process happens before each iteration begins, we fuse the SREncode with the update process (optimizer step) of the last iteration for less overhead.
In \textit{Asyn-comm} stage, the Send Queue sequentially pop expert residuals for AG communication.
The Recv Queue receives the corresponding results and send to SRDecode for the subsequent expert computation.
Note that this communication process is parallel to the pre-expert computation process of the model so we can overlap them.
Moreover, we fused the SRDecode with expert computation for better efficiency.

\section{Evaluation}
\label{experiment}

Our experiments aim to answer the following questions:

\begin{itemize}[itemsep=2pt,topsep=0pt,parsep=0pt,leftmargin=10pt]
    \item Is our stream-based modeling accurately estimating computation, communication, and determining the best proportion with minimal latency? (\S\ref{modeling-effectiveness})

    \item What is the end-to-end speedup of HybridEP with different data/expert size? (\S\ref{overall-speedup})

    \item How much does the domain-based partition and parameter-efficient migration contribute to the final effect? (\S\ref{ablation-study})

    \item Does efficient parameter migration affect training accuracy and what's the impact of its compression/decompression process on computation? (\S\ref{model-performance})

    \item Does HybridEP has better communication traffic and frequency characteristics compared with EP? (\S\ref{sec:comm-characteristic})

    % \item Will HybridEP still be effective on a larger scale? (\S\ref{sec:simulation})
    \item As a more general framework, does HybridEP have better scalability than EP on larger scale? (\S\ref{sec:simulation})
    
\end{itemize}

\subsection{Experiment Setup}

\textbf{Testbed.}
We conduct experiments on three clusters consisting of different number of DCs. 
Due to the limitations of the actual environment, we regard a single node as a DC, which is internally connected by PCIe3.0 x16 (128 Gbps), and DCs are connected by a low bandwidth of Ethernet (10 Gbps).
Specifically, we have
\ding{172} \textbf{Cluster-S:} a cluster with 8 $\times$ NVIDIA A800 GPUs in a single DC.
\ding{173} \textbf{Cluster-M:} a cluster with 16 $\times$ NVIDIA A800 GPUs on 2 DCs. 
\ding{174} \textbf{Cluster-L:} a cluster with 32 $\times$ NVIDIA A800 GPUs on 4 DCs.
% \ding{175} \textbf{Cluster-L:} a cluster with 32 $\times$ NVIDIA A800 GPUs on 8 nodes.
% Moreover, GPUs in the same node are connected through PCIe3.0 x16 (128 Gbps), and GPUs in different nodes are connected through Ethernet (10 Gbps). 
Note that we use Cluster-S to verify the effectiveness of our modeling without considering hierarchical architecture, while using Cluster-M and Cluster-L to verify the effectiveness of HybridEP in real-world training tasks.
Moreover, HybridEP is built based on Tutel~\cite{tutel} and Pytorch v.1.12.1, and the experiment environment is under Ubuntu-18.04, CUDA-11.3, cuDNN-7.6, and NCCL-2.10.

\begin{table}[t]
\setlength{\abovecaptionskip}{0.5em}
    \setlength{\belowcaptionskip}{0.5em}
\caption{\textbf{Configurations of Models.}
% \emph{$E$ is the expert number, $H$ is the hidden size, $P_E$ is each expert's parameter number, and Layer is the total number of (MoE and Transformer) blocks.}
}
\begin{center}
\small
\resizebox{\linewidth}{!}{
\begin{tabular}{c|c|c|c|c|c}
\hline
\textbf{Model}& \textbf{Dataset} & \textbf{E}&
\textbf{H}&$\mathbf{P_E}$&\textbf{\#Layers}\\
\hline\hline
Llama-Tiny & PennTreebank~\cite{Penn-Treebank} & 32 & 512 & 2.1M &$12$\\
Mistral-Small & WikiText2~\cite{wikitext} & 32 & 768 & 4.7M & $12$\\
GPT-Medium & OpenWebText-10k~\cite{openwebtext-10k} & 32 & 1024 & 8.4M & $12$\\
GPT-Large & WikiText103~\cite{wikitext}  & 32 & 1024 & 8.4M & $16$\\
\hline
\end{tabular}
}
\label{tab:model-config}
\vspace{-1em}
\end{center}
\end{table}

\textbf{Configurations of Models, Datasets, and Compared Methods.}
We summarize tested models and datasets in Table~\ref{tab:model-config}. Specifically, 
\ding{172} We use Llama-Tiny~\cite{llama3} for PennTreebank~\cite{Penn-Treebank} dataset, which is one of the most known and used corpus for the evaluation of models for sequence labeling.
\ding{173} We use Mistral-Small~\cite{mistral-moe} for wikitext2~\cite{wikitext} dataset, which is a collection of over 100 million tokens extracted from the set of verified Good and Featured articles on Wikipedia;
\ding{174} We use GPT-Medium~\cite{gpt} for OpenWebText-10k, which is an open-source replication of the WebText dataset from OpenAI~\cite{openwebtext-10k};
\ding{175} We use GPT-Large~\cite{gpt} for WikiText103~\cite{wikitext}, which is similar to wikitext2 but much larger.
Note that we only built a smaller version for training based on the above model structure, not the original one.
We compare HybridEP with Tutel~\cite{tutel}, FasterMoE~\cite{fastermoe} and SmartMoE~\cite{smartmoe}.
These MoE-specific optimized systems focus on dimensions of data transmission, expert transmission, and pipeline, which are commonly used in HPC environment. 
% therefore comparing to them can reflect the effectiveness of HybridEP.
Note that we do not compare to some training systems~\cite{megatron-lm, deepspeedmoe} because they also make some other optimizations besides MoE, which is also adopted by many works~\cite{fastermoe,smartmoe,hetumoe,schemoe, scomoe}.

\begin{table}[t]
    \setlength{\abovecaptionskip}{0.5em}
    \setlength{\belowcaptionskip}{0.5em}
    \centering
    \caption{\textbf{Configurations of MoE Layers.} 
    }
    \small
    \begin{tabular}{c|l}
    \hline
    \textbf{Parameter} & \textbf{Candidate Values} \\ \hline\hline
    $K$ & $\{1, 2, 4\}$ \\
    $B$ & $\{8, 16, 32\}$ \\ 
    $L$ & $\{128, 256, 512\}$ \\ 
    $H$ & $\{512, 768, 1024\}$ \\ 
    $M$ & $\{768,1024,1536,2048,3072,4096\}$ \\ \hline
    \end{tabular}
    \label{tab:hyperparameter}
\end{table}

\textbf{Extra Configurations.}
We use Adam optimizer for all experiments with a learning rate of $1e-4$ and Pytorch DDP for backward propagation, which can efficiently synchronize gradients of model parameters using ll-Reduce.
Note that we do not use Zero Optimizer\cite{zero} for the non-MoE part and also the pipeline parallelism due to the potential network bandwidth conflicts, which may affect our model's accuracy.
Moreover, all configurations will be adjusted within Table~\ref{tab:hyperparameter} to meet different experiment requirements.
Specifically, $K$ is the number of activated experts, $B$ is the batch size, $L$ is the sequence length, and $H, M$ are experts' two dimensions.

\subsection{Modeling Verification}
\label{modeling-effectiveness}

% \subsection{The Effectiveness of Our Modeling}
% \label{modeling-effectiveness}

To verify modeling effectiveness, \ding{172} we first verify whether it can accurately estimate the computation and communication latency, \ding{173} we then verify whether it can find the optimal proportion of A2A and AG ($p$ in Figure~\ref{function}). 
% Note that all experiments are conducted on Cluster-S with one node due to the consumption of homogeneous bandwidth. 
% However, it is still effective under heterogeneous bandwidth thanks to multi-level scaling (Figure~\ref{scaling-verify}).

\begin{figure}[t]
    \setlength{\abovecaptionskip}{0.5em}
    \setlength{\belowcaptionskip}{0.5em}
    \centering
    \includegraphics[width=\linewidth]{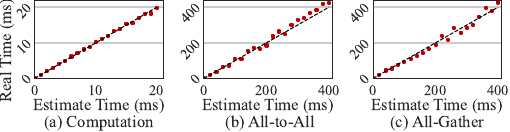}
    \vspace{-0.5cm}
    \caption{\textbf{Latency Verification of Comp. and Comm..} \emph{Since the estimate computation, A2A, AG latency (red markers) are close to the real latency (black line), our stream-based modeling can effectively model system latency.}}
    \label{FIG-Modeling-Effectiveness}
    \vspace{-1em}
\end{figure}

\textbf{Verification of Estimated Computation and Communication Latency.}
We adjust the sizes of the data traffic and expert size to test the accuracy of our model, as shown in Figure~\ref{FIG-Modeling-Effectiveness}.
% test whether our modeling can accurately estimate the computation, A2A, and AG latency, as shown in Figure~\ref{FIG-Modeling-Effectiveness}. 
Results suggest that the estimated latency is close to the real latency. However, they are fluctuating because our experiment platform is shared by multiple users with unstable network bandwidth. Nevertheless, such small fluctuations do not affect the effectiveness of our model.

\begin{table}[t]
\setlength{\abovecaptionskip}{0.5em}
\setlength{\belowcaptionskip}{0.5em}
\caption{\textbf{Configurations of Modeling Verification.} 
}
\resizebox{\linewidth}{!}{
\begin{tabular}{c|c|c|c|c|c|c}
\hline
\textbf{Case} & $\bm p$ & $\bm G$ & $\bm B$ & $\bm Lat_{comp}^{PE}$ & $\bm D$ & $\bm P_E$\\ \hline\hline
Mix-1   & 0.75 & 8 & 128 Gbps & 0.049 ms  & 8 MB & 4.7 MB \\ 
Mix-2 & 0.5 & 8 & 128 Gbps  & 0.049 ms & 8 MB & 2.35 MB \\
AG-only-1 & 0   & 8    & 128 Gbps & 0.099 ms & 3 MB & 0.094 MB \\ 
AG-only-2 & 0   & 8  & 128 Gbps  & 0.099 ms & 3 MB & 0.047 MB \\ \hline
\end{tabular}
}
\label{tab:modeling-effectiveness-config}
\vspace{-1em}
\end{table}

\begin{figure}[t]
\setlength{\abovecaptionskip}{0.5em}
    \setlength{\belowcaptionskip}{0.5em}
    \centering
    \includegraphics[width=\linewidth]{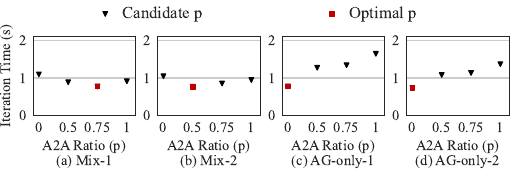}
    \vspace{-0.5cm}
    \caption{\textbf{Modeling Verification.} \emph{Results suggest that our modeling can find the optimal $p$ (red marker) with the least iteration time among candidate configurations (black marker).}}
    \label{FIG-Optimal-p}
    \vspace{-1em}
\end{figure}

\textbf{Verification of the Optimal $p$.} We then adjust the training configurations to verify whether our modeling can find the optimal proportion $p$ of A2A and AG in different cases, as shown in Table~\ref{tab:modeling-effectiveness-config}. 
Note that one node has 8 GPUs in our configuration. Therefore, the candidates $p$ are $0, 0.5, 0.75, 1$, which indicates that the expert domain size is $8, 4, 2, 1$, respectively.
The results are shown in Figure~\ref{FIG-Optimal-p}, where the optimal $p$ has the lowest average iteration latency among 4 candidate $p$, demonstrating the effectiveness of our model. Specifically, Mix-1 and Mix-2 represent Case 2.1 in Figure~\ref{function} (i.e., $2D-GP_E<0$), therefore HybridEP communicates through both A2A and AG, and our modeling finds the optimal proportion of A2A data (i.e., $p = 0.5, 0.25$). 
Moreover, AG-only-1 and AG-only-2 represent Case 2.2 in Figure~\ref{function} (i.e., $2D-GP_E\geq0$), therefore HybridEP should communicate only through AG (i.e., $p = 0$) for the lowest iteration latency.

\begin{table*}[thbp]
% \centering
\setlength{\abovecaptionskip}{0.5em}
\setlength{\belowcaptionskip}{0.5em}
\caption{\textbf{Average Iteration Time (in Seconds) and Average Speedup ($\times$) under Different Data Traffic Sizes.} 
}\label{tab:speedup-data-size}
\resizebox{\linewidth}{!}{
    \begin{tabular}{c|rrrrrr|rrrrrr}
    \hline
    \multirow{2}{*}{\diagbox{\textbf{Method}}{\textbf{Data}}} & \multicolumn{6}{c|}{\textbf{Cluster-M}}                                                                                                                                                                                                                                                                                                                                                                                                                                                           & \multicolumn{6}{c}{\textbf{Cluster-L}}                                                                                                                                                                                                                                                                                                                                                                                                                                                              \\ \cline{2-13} 
                                      & \multicolumn{1}{c|}{\textbf{6 MB}}                                                  & \multicolumn{1}{c|}{\textbf{12 MB}}                                                 & \multicolumn{1}{c|}{\textbf{24 MB}}                                                 & \multicolumn{1}{c|}{\textbf{48 MB}}                                                & \multicolumn{1}{c|}{\textbf{96 MB}}                                                & \multicolumn{1}{c|}{\textbf{192 MB}}                           & \multicolumn{1}{c|}{\textbf{6 MB}}                                                  & \multicolumn{1}{c|}{\textbf{12 MB}}                                                 & \multicolumn{1}{c|}{\textbf{24 MB}}                                                 & \multicolumn{1}{c|}{\textbf{48 MB}}                                                 & \multicolumn{1}{c|}{\textbf{96 MB}}                                                 & \multicolumn{1}{c}{\textbf{192 MB}}                            \\ \hline\hline
    Tutel                             & \multicolumn{1}{r|}{2.52 s}                                                           & \multicolumn{1}{r|}{4.26 s}                                                           & \multicolumn{1}{r|}{5.82 s}                                                           & \multicolumn{1}{r|}{7.62 s}                                                          & \multicolumn{1}{r|}{12.65 s}                                                         & 20.35 s                                                          & \multicolumn{1}{r|}{3.74 s}                                                           & \multicolumn{1}{r|}{7.30 s}                                                            & \multicolumn{1}{r|}{10.69 s}                                                          & \multicolumn{1}{r|}{13.54 s}                                                          & \multicolumn{1}{r|}{18.59 s}                                                          & 28.46 s                                                          \\
    FasterMoE                         & \multicolumn{1}{r|}{2.58 s}                                                           & \multicolumn{1}{r|}{4.37 s}                                                           & \multicolumn{1}{r|}{5.90 s}                                                            & \multicolumn{1}{r|}{7.81 s}                                                          & \multicolumn{1}{r|}{12.80 s}                                                          & 20.82 s                                                          & \multicolumn{1}{r|}{3.86 s}                                                           & \multicolumn{1}{r|}{7.50 s}                                                            & \multicolumn{1}{r|}{11.09 s}                                                          & \multicolumn{1}{r|}{13.88 s}                                                          & \multicolumn{1}{r|}{19.32 s}                                                          & 29.43 s                                                          \\
    SmartMoE                          & \multicolumn{1}{r|}{2.59 s}                                                           & \multicolumn{1}{r|}{4.34 s}                                                           & \multicolumn{1}{r|}{5.97 s}                                                           & \multicolumn{1}{r|}{7.80 s}                                                           & \multicolumn{1}{r|}{12.68 s}                                                         & 20.91 s                                                          & \multicolumn{1}{r|}{3.82 s}                                                           & \multicolumn{1}{r|}{7.46 s}                                                           & \multicolumn{1}{r|}{10.94 s}                                                          & \multicolumn{1}{r|}{14.08 s}                                                          & \multicolumn{1}{r|}{19.25 s}                                                          & 29.53 s                                                          \\
    HybridEP (Ours)                    & \multicolumn{1}{r|}{2.48 s}                                                           & \multicolumn{1}{r|}{2.63 s}                                                           & \multicolumn{1}{r|}{2.74 s}                                                           & \multicolumn{1}{r|}{2.82 s}                                                          & \multicolumn{1}{r|}{3.01 s}                                                          & 3.78 s                                                           & \multicolumn{1}{r|}{3.49 s}                                                           & \multicolumn{1}{r|}{3.53 s}                                                           & \multicolumn{1}{r|}{3.54 s}                                                           & \multicolumn{1}{r|}{3.85 s}                                                           & \multicolumn{1}{r|}{4.24 s}                                                           & 5.20 s                                                            \\ \hline\hline
    \textbf{Avg. Speedup}                  & \multicolumn{1}{r|}{\textbf{\begin{tabular}[c]{@{}r@{}}1.03$\times$\end{tabular}}} & \multicolumn{1}{r|}{\textbf{\begin{tabular}[c]{@{}r@{}}1.64$\times$\end{tabular}}} & \multicolumn{1}{r|}{\textbf{\begin{tabular}[c]{@{}r@{}}2.15$\times$\end{tabular}}} & \multicolumn{1}{r|}{\textbf{\begin{tabular}[c]{@{}r@{}}2.75$\times$\end{tabular}}} & \multicolumn{1}{r|}{\textbf{\begin{tabular}[c]{@{}r@{}}4.22$\times$\end{tabular}}} & \multicolumn{1}{r|}{\textbf{\begin{tabular}[c]{@{}r@{}}5.47$\times$\end{tabular}}} & \multicolumn{1}{r|}{\textbf{\begin{tabular}[c]{@{}r@{}}1.09$\times$\end{tabular}}} & \multicolumn{1}{r|}{\textbf{\begin{tabular}[c]{@{}r@{}}2.10$\times$\end{tabular}}} & \multicolumn{1}{r|}{\textbf{\begin{tabular}[c]{@{}r@{}}3.08$\times$\end{tabular}}} & \multicolumn{1}{r|}{\textbf{\begin{tabular}[c]{@{}r@{}}3.59$\times$\end{tabular}}} & \multicolumn{1}{r|}{\textbf{\begin{tabular}[c]{@{}r@{}}4.49$\times$\end{tabular}}} & \textbf{\begin{tabular}[c]{@{}r@{}}5.60$\times$\end{tabular}} \\ \hline
    \end{tabular}
}
\vspace{-2em}
\end{table*}

\subsection{End-to-end Speedup}
\label{overall-speedup}
We test HybridEP in Cluster-M and Cluster-L with different MoE configurations (Table~\ref{tab:hyperparameter}) in
two scenarios.
% communication-dominated scenarios. 
Specifically, \ding{172} different data traffic ranging from 6 MB to 192 MB; \ding{173} different expert size ranging from 32 MB to 2 MB.

\textbf{Different Data Traffic.} 
We change data traffic from 6 to 192 MB 
% (under 10 Gbps Ethernet and 128 Gbps PCIe) 
and fix expert size to 0.36 MB.
The results are shown in Table~\ref{tab:speedup-data-size}, where HybridEP achieves an average speedup of up to 5.60 $\times$. 
Specifically, with larger data traffic, lower bandwidth, and more connected DCs, 
The communication bottleneck of EP becomes more and more obvious.
% communication gradually becomes the bottleneck. 
However, HybridEP finds the appropriate proportion of A2A and AG ($p$ in Figure~\ref{function}), thus achieving significant speedup.

\begin{figure}[t]
\setlength{\abovecaptionskip}{0.5em}
    \setlength{\belowcaptionskip}{0.5em}
    \centering
    \includegraphics[width=\linewidth]{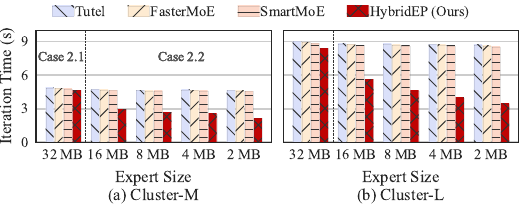}
    \vspace{-0.5cm}
    \caption{\textbf{Average Iteration Time under Different Expert Sizes.} \emph{Results suggest that as the expert size decreases, the computation cost decreases, HybridEP's iteration latency decreases. However, iteration latency of compared methods is nearly unchanged, despite the decreased computation overhead (i.e., $\frac{1}{16}$, expert size decreases from 32 MB to 2 MB). 
    % Moreover, HybridEP has little speedup in Case 2.1 (Figure~\ref{function}), where computation becomes the main bottleneck.
    }}
    \label{FIG-Expert-Size}
    \vspace{-1em}
\end{figure}

\textbf{Different Expert Size.} 
We change expert size from 32 MB to 2 MB and fix the data traffic to 16 MB. Therefore, computation cost decreases as expert sizes decrease, and we do not use the SR expert compression for better observation.
% We verify the speedup of HybridEP by reducing the expert size to aggravate the communication bottleneck with fixed data volume 16MB (without our compressor).
% In this test, we fix the data volume size to 16 MB (without compression), then change the expert size to show how computation affects training latency.
The results are shown in Figure~\ref{FIG-Expert-Size}, where HybridEP achieves a speedup ranging from 1.18 $\times$ to 2.57 $\times$. Specifically, as the expert size decreases, 
HybridEP can transmit more experts with small traffic, thus enlarging the expert domain size and reducing EP's overhead.
% the computation cost is decreased, and HybridEP's average iteration latency is also decreased. 
% However, compared methods have little speedup because A2A is always the bottleneck.
Thus, the acceleration effect of case 2.1 is not as significant as that of case 2.2. 
% This is because the large expert size leads to smaller expert domain and gradually shifts the bottleneck to the computation.
However, Case 2.1 can be transformed into Case 2.2 for higher speed with SR compression to change the condition to $2D - GP_E \geq 0$.

\begin{table}[t]
\setlength{\abovecaptionskip}{0.5em}
\setlength{\belowcaptionskip}{0.5em}
% \caption{\textbf{Ablation Study.} \emph{The tested data volume size is 24 MB, while the tested expert parameter size is 8 MB. Moreover, + Scaling is untested in Cluster-S (i.e., '--') because it has only one node with homogeneous and high-speed bandwidth.}}
\caption{
\textbf{Ablation Study.}
% \emph{Data\&Expert represents the data traffic and expert size.
% Partition represents only using the domain-based partition to implement the modeling, while + Migration represents adding parameter-efficient migration to the partition, that is HybridEP.}
}
\label{tab:ablation-study}
\small
\centering
\begin{tabular}{c|r|r|r}
\hline
\textbf{Cluster} & \multicolumn{1}{c|}{\textbf{Data\&Expert}} & \multicolumn{1}{c|}{\textbf{Partition}} & \multicolumn{1}{c}{\textbf{+ Migration}}
% & \multicolumn{1}{c}{\textbf{HybridEP}}
\\ \hline\hline
Cluster--S    &24\&8 MB                           & 0.76 s                       & 0.61 s                              \\
Cluster--M     & 24\&8 MB                           & 3.41 s                            & 2.54 s                           \\
Cluster--L     & 24\&8 MB                           & 6.12 s                            & 3.48 s                           \\ \hline
Cluster--S    &48\&2 MB                           & 1.06 s                       & 0.74 s                              \\
Cluster--M     & 48\&2 MB                           & 6.21 s                            & 2.81 s                           \\
Cluster--L     & 48\&2 MB                           & 10.89 s                            & 3.86 s                           \\ \hline
\end{tabular}
\vspace{-0.1cm}
\end{table}

\subsection{Ablation Study}
\label{ablation-study}

In this section, we evaluate how domain-based partition (baseline) and parameter-efficient migration contributes to the overall speedup with different data traffic and expert size.

\textbf{Configurations and Results Analysis.} 
In Table~\ref{tab:ablation-study}, Data\&Expert represent the size of data and expert.
The remaining two items correspond to the two designs of HybridEP (\emph{+Migration} equals to HybridEP).
For 24\&8MB configuration,
our modeling suggests that Cluster-S has $p=0.5$ (i.e., $S_{ED}^0=4$), while Cluster-M and Cluster-L has two levels, denote as $S_{ED}^0=2,S_{ED}^1=2$ and $S_{ED}^0=4,S_{ED}^1=1$. 
For 48\&2MB configuration, $p$ is 0 for all clusters.
\emph{+Migration} adds parameter-efficient migration (i.e., HybridEP).
Table~\ref{tab:ablation-study} suggests that
\emph{+Migration} (i.e., HybridEP)
% \emph{+ Scaling} and \emph{+ Compr.} 
achieves a speedup of 1.25$\times$ to
2.82$\times$,
compared to the baseline \emph{Partition}.
Larger data traffic and smaller expert size contribute to faster training speed.
Note that the $S_{ED}$ in our experiments includes all DCs, so the more DCs that are interconnected, the more significant the speedup. However, this may not be always true in practice, more details in \S\ref{sec:simulation}.

\begin{figure}[t]
\setlength{\abovecaptionskip}{0.5em}
    \setlength{\belowcaptionskip}{0.5em}
    \centering
    \includegraphics[width=\linewidth]{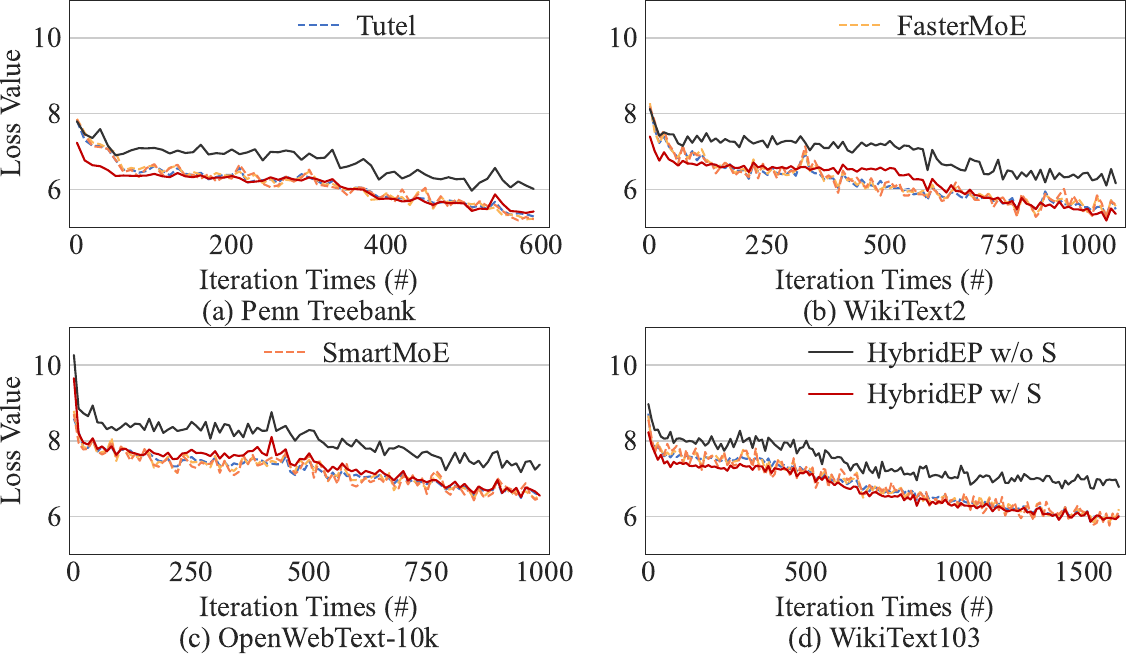}
    \vspace{-0.5cm}
    \caption{\textbf{Loss Analysis}. \emph{HybridEP's expert compression ratio is 50 $\times$. Moreover, w/o S indicates that HybridEP directly compress experts, while w/ S indicates that HybridEP compress experts with shared expert (\S\ref{efficient-comm}).}}
    \label{FIG-Accuracy}
    \vspace{-1em}
\end{figure}

\begin{figure}[t]
\setlength{\abovecaptionskip}{0.5em}
    \setlength{\belowcaptionskip}{0.5em}
    \centering
    \includegraphics[width=\linewidth]{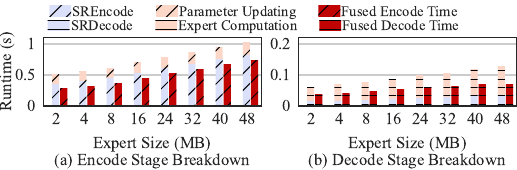}
    \vspace{-0.5cm}
    \caption{
    \textbf{Time breakdown of Parameter-Efficient Migration's Tow Phases.}
    \emph{Under different expert sizes, (a) shows the effect of SREncode fused with the parameter update of last iteration, which can reduce overhead by 30\%. 
    (b) shows the effect of SRDecode fused with multiple expert computations, which can reduce overhead by 45\%.}
    } 
    \label{time-breakdown}
    % \vspace{-1em}
\end{figure}

\subsection{Analysis on Migration}
\label{model-performance}

In this section, we first evaluate whether the SR expert compressor affects accuracy.
Then we conduct the time breakdown experiments to demonstrate how the two phases of SR compression fused with other operations for less overhead.
% to reduce communication overhead while adding negligible computational overhead. 
Our configurations are shown in Table~\ref{tab:model-config}.
% Specifically, we test HybridEP in tasks Penn Treebank, WikiText2, OpenWebText-10k, and WikiText103 and 
% We record the loss of different iterations.

\textbf{Configurations of SR-Based Expert Compression.} 
HybridEP has a hyperparameter (compression ratio, CR) to control SR-based expert compression considering model accuracy and compressibility.
% , with considerations of model accuracy (i.e., lower loss) and compressibility. 
% We set CR = 50 in all of our experiments 
We use \emph{HybridEP w/ S} to represent the compression with shared expert, and use \emph{HybridEP w/o S} to represent the naive method that directly compresses the expert through Top-k.
Our goal is to find the maximum CR without affecting model performance.

\textbf{Results of Accuracy Analysis.}
Figure~\ref{FIG-Accuracy} suggests that the loss value of \emph{HybridEP w/ S} is close to that of the compared methods (i.e., Tutel, FasterMoE, and SmartMoE). Therefore, our proposed SR-based compression algorithm can retain both a high compression ratio (i.e., 50 $\times$, we do not display other results due to the page limit) and high accuracy. In contrast, \emph{HybridEP w/o S}'s loss value is quite higher than compared methods, which indicates that the shared expert in our design plays an important role in accuracy maintenance.

\textbf{Time Breakdown Analysis.}
As shown in Figure~\ref{time-breakdown}, as the expert size increases, the time overhead of both SREncode and SRDecode increases. 
When integrated with other computations, the overheads can be further reduced by up to 30\% and 45\%, respectively. 
Although they are not completely eliminated, it is not significant compared to the communication and remains within acceptable limits.
However, designing more efficient expert compression is still worth exploring.

\begin{figure}[t]
\setlength{\abovecaptionskip}{0.5em}
    \setlength{\belowcaptionskip}{0.5em}
    \centering
    \includegraphics[width=\linewidth]{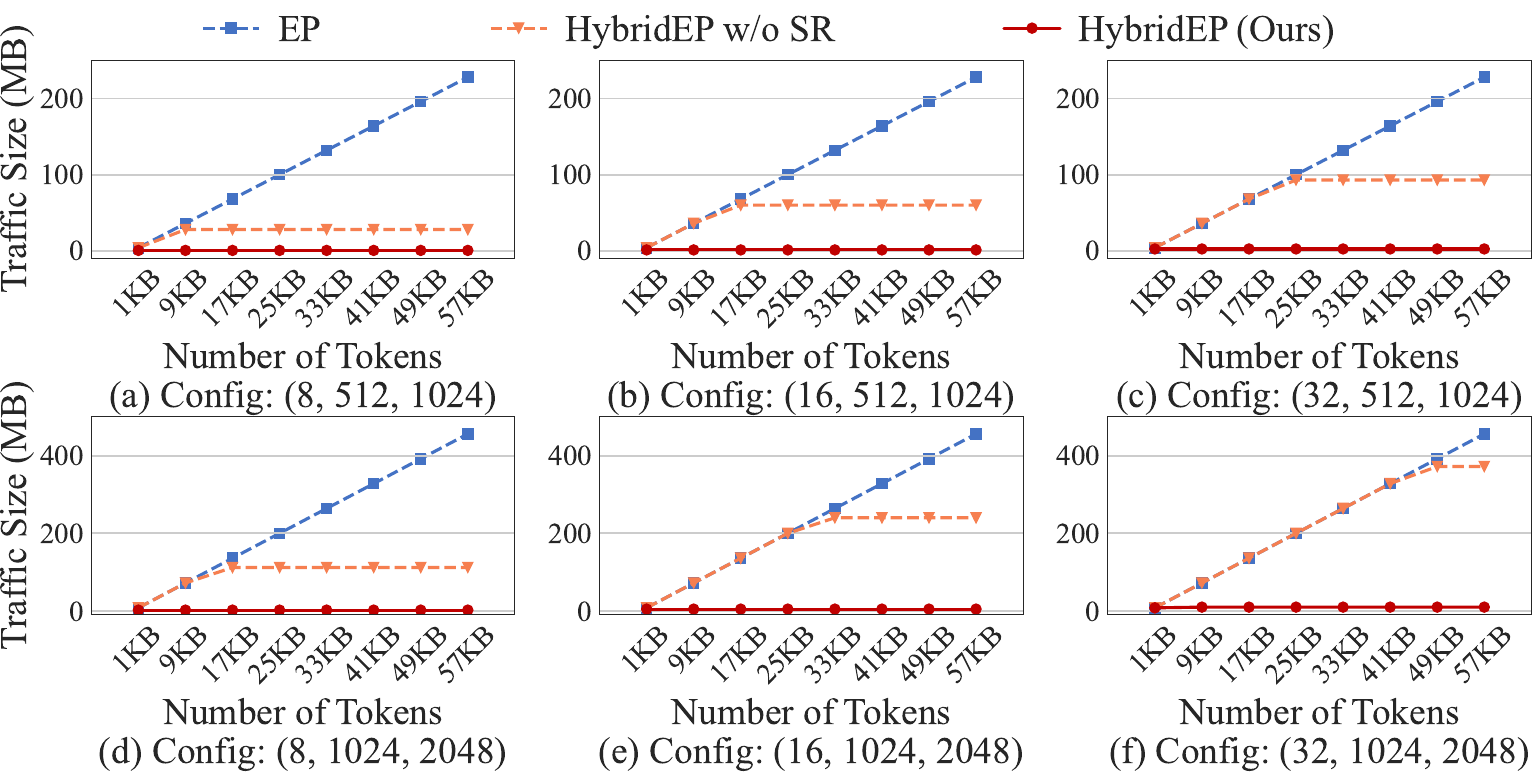}
    \vspace{-0.5cm}
    \caption{\textbf{Traffic Scalability 
 Analysis}. \emph{HybridEP has less communication traffic under constrained bandwidth, leading to a better scalability.
 The configuration is a triplet, representing the size of EP and the tow dimensions of expert weights ($H$ \& $M$).}}
    \label{fig:traffic}
    \vspace{-1em}
\end{figure}

\begin{table}[t]
\caption{\textbf{Communication Frequency with Different EP Size.}
% \emph{As the domain size increases, the frequency of A2A communication in EP gradually decreases, and the frequency of AG communication gradually increases instead.
% Note that expert domain size can not exceed the EP size.}
}
\centering
\begin{tabular}{cc|cccccc}
\hline
\multicolumn{1}{c|}{\multirow{2}{*}{\textbf{\begin{tabular}[c]{@{}c@{}}EP\\Size\end{tabular}}}} & \multicolumn{1}{c|}{\multirow{2}{*}{\textbf{\begin{tabular}[c]{@{}c@{}}Comm.\\Type\end{tabular}}}} & \multicolumn{6}{c}{\textbf{Expert Domain Size ($S_{ED}$)}}                                                                                                                                               \\ \cline{3-8} 
\multicolumn{1}{c|}{} & \multicolumn{1}{c|}{}                                                                                   & \multicolumn{1}{c|}{\textbf{1 (EP)}} & \multicolumn{1}{c|}{\textbf{2}} & \multicolumn{1}{c|}{\textbf{4}} & \multicolumn{1}{c|}{\textbf{8}} & \multicolumn{1}{c|}{\textbf{16}} & \textbf{32} \\ \hline\hline
\multicolumn{1}{c|}{\multirow{2}{*}{8}}                               & A2A                             & \multicolumn{1}{c|}{56}         & \multicolumn{1}{c|}{24}         & \multicolumn{1}{c|}{8}          & \multicolumn{1}{c|}{0}          & \multicolumn{1}{c|}{-}           & -           \\
\multicolumn{1}{c|}{}                                                 & AG                              & \multicolumn{1}{c|}{0}          & \multicolumn{1}{c|}{8}          & \multicolumn{1}{c|}{24}         & \multicolumn{1}{c|}{56}         & \multicolumn{1}{c|}{-}           & -           \\ \hline
\multicolumn{1}{c|}{\multirow{2}{*}{16}}                              & A2A                             & \multicolumn{1}{c|}{240}        & \multicolumn{1}{c|}{112}        & \multicolumn{1}{c|}{48}         & \multicolumn{1}{c|}{16}         & \multicolumn{1}{c|}{0}           & -           \\
\multicolumn{1}{c|}{}                                                 & AG                              & \multicolumn{1}{c|}{0}          & \multicolumn{1}{c|}{16}         & \multicolumn{1}{c|}{48}         & \multicolumn{1}{c|}{112}        & \multicolumn{1}{c|}{240}         & -           \\ \hline
\multicolumn{1}{c|}{\multirow{2}{*}{32}}                              & A2A                             & \multicolumn{1}{c|}{992}        & \multicolumn{1}{c|}{480}        & \multicolumn{1}{c|}{224}        & \multicolumn{1}{c|}{96}         & \multicolumn{1}{c|}{32}          & 0           \\
\multicolumn{1}{c|}{}                                                 & AG                              & \multicolumn{1}{c|}{0}          & \multicolumn{1}{c|}{32}         & \multicolumn{1}{c|}{96}         & \multicolumn{1}{c|}{224}        & \multicolumn{1}{c|}{480}         & 992         \\ \hline
\label{tab:frequency-compare}
\end{tabular}
\vspace{-0.4cm}
\end{table}

\subsection{EP vs. HybridEP: Characteristic Comparison}
\label{sec:comm-characteristic}
In this section, we show the comparison between HybridEP and EP in terms of communication traffic and frequency under different configurations.

\textbf{Traffic Analysis.}
As shown in Figure~\ref{fig:traffic}, the traffic of original EP grows linearly with the number of tokens during each training iteration.
In contrast, HybridEP introduces a more fixed and input-independent traffic with limited upper bound.
% once an expert is transferred to the target device, it can serve multiple tokens locally, regardless of how many tokens choose that expert.
When the number of tokens is small, HybridEP's traffic is almost the same as EP's. 
However, when the number of tokens increases significantly, EP becomes a huge communication bottleneck, while HybridEP guarantees a fixed traffic via only transmitting experts.
This makes HybridEP more predictable and stable, which is especially advantageous in low-bandwidth or burst-sensitive environments.

\textbf{Frequency Analysis.}
We use the sum of all GPU-to-GPU communications as frequency. 
The comparison is shown in Table~\ref{tab:frequency-compare}. 
Note that $S_{ED}=1$ represents the original EP.
As the expert domain expands, the A2A communication frequency decreases quadratically, while the AG frequency increases accordingly. 
This can be seen as a gradual shift of A2A communication to AG. However, due to the more asynchronous nature of AG and its ability to significantly reduce traffic via compression, HybridEP achieves higher efficiency.

\begin{figure}[t]
    \centering
    % 第一个子图
    \begin{minipage}[b]{\linewidth} % 第一个子图区域，宽度可以更宽
        \centering
        \includegraphics[width=\linewidth]{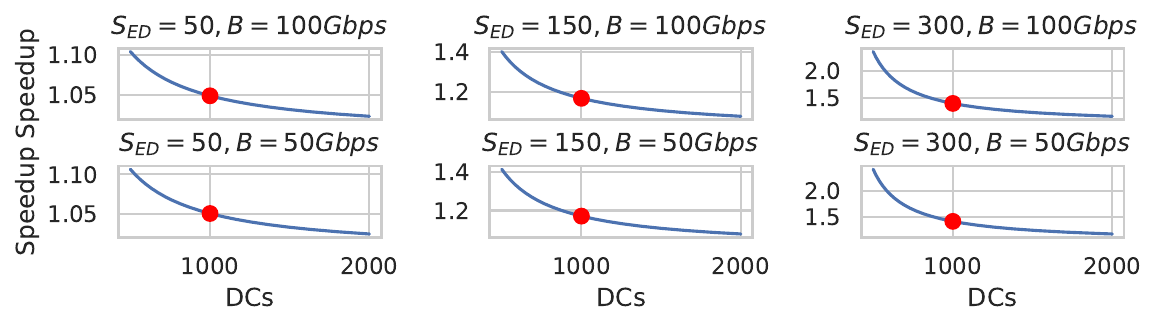}
        \vspace{0.5\baselineskip}
        \small (a) Fixed $S_{ED}$ and dynamic $p$. \\
        \label{fig:subim1}
    \end{minipage}
    \par % 强制换行
    % \vspace{4cm} % 可选：增加垂直间距

    % 第二个子图
    \begin{minipage}[b]{\linewidth} % 第二个子图区域
        \centering
        \includegraphics[width=\linewidth]{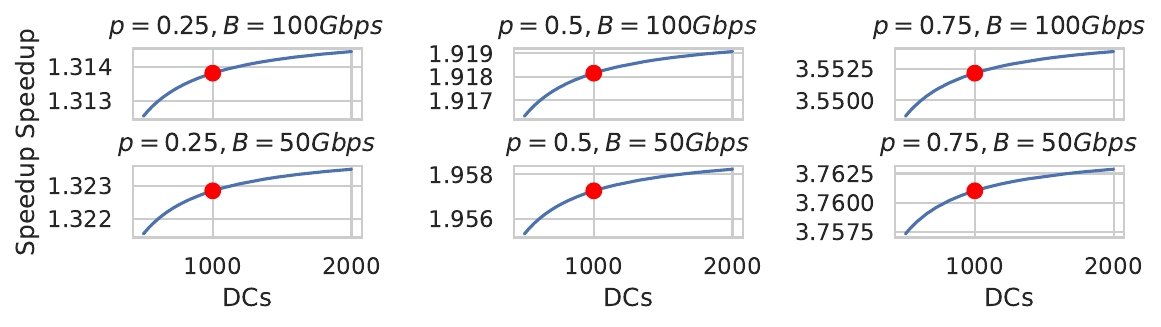}
        \vspace{0.5\baselineskip}
        \small (b) Fixed $p$ and dynamic $S_{ED}$.  \\
        \label{fig:subim2}
    \end{minipage}
    \vspace{-0.8cm}
    \caption{\textbf{Speedup of HybridEP on Large Scale Simulation.}}
    \label{fig:simulation}
    \vspace{-1em}
\end{figure}

\subsection{EP vs. HybridEP: Large Scale Simulation}
\label{sec:simulation}
In this section, we conduct the performance simulation with SimAI\cite{simAI} due to limited environments.
We test on a large scale to verify HybridEP's effectiveness under general settings.

\textbf{Results of Simulation.}
We evaluate the effectiveness under different bandwidth from two cases, as shown in Figure~\ref{fig:simulation}. 
We first fix the expert domain size $S_{ED}$ and expand the number of DCs. 
This essentially reduces the proportion $p$ determined by HybridEP's model, resulting in a smaller acceleration effect.
Given 1000 DCs (red dot), HybridEP achieves a speedup of 1.05$\times$ to 1.45$\times$. 
Then, we fix the proportion $p$ and expand the number of DCs. 
This essentially increases the size of domain, resulting in an improvement in the acceleration effect. 
Given 1000 DCs, HybridEP achieves 1.31$\times$ to 3.76$\times$ speedup.
Furthermore, in both cases, the lower the bandwidth, the greater the speedup. 
% although the difference is not significant.
% However, the range of improvement is very narrow, so it can be roughly considered unchanged. 
Note that in practice, the first case is the most common because $S_{ED}$ is fixed due to the fixed training configurations. 
Thus, the speedup decreases as the number of DCs increases.
How to expand in the second case still remains extremely challenging, which is widely recognized.
\section{Discussion}

\textbf{Storage Overhead of SR-Based Expert Compression.}
The additional storage overhead introduced by our proposed expert compression algorithm can be handled. Specifically, it consists of expert residual and shared expert.
\ding{172} The expert residual consumes little GPU memory due to its high compressibility.
\ding{173} The shared experts compete with local experts for GPU memory, which can be solved by offloading local experts to CPU memory while keeping shared experts in GPU memory. 
% The rationale is that, in a large cluster, most data are processed by shared experts (i.e., in the original MoE system, most data will be sent to other GPUs through A2A), while only a few data are processed by local experts. 
Offloading local experts to CPU memory is an effective strategy, which has been well studied (e.g., Zero-Offload~\cite{zero-offload}) and can be directly integrated into HybridEP.

\textbf{Backward Propagation in MoE Training.} The backward phase has the unique All-Reduce communication to synchronize model parameters, which competes with other types of communications and thus affects our modeling. Nevertheless, our modeling is still effective for backward propagation because the All-Reduce communication traffic is relative to the model size, and its latency can be regarded as a constant when model configurations are determined. Therefore, our modeling can handle backward propagation by simply adding a constant.

\section{Related Works}

% In recent years, lots of works~\cite{horovod, megatron-lm,fairseq, alpa, flexflow} have focused on the optimization of distributed training.
% Since the sparsely activated MoE~\cite{moe} has been successfully applied in Transformers~\cite{GShard, vmoe}, some optimizations came out to improve MoE training performance~\cite{SEmoe, fastmoe, tutel, deepspeedmoe},which can be grouped into three main directions.

% To the best of our knowledge, we are the first work to scale the EP under build a stream-based MoE modeling with considerations of lossy compression, pipeline, and prefetch. 
We introduce works that are orthogonal (or related) to our study, which mainly focus within the high performance cluster.

\textbf{Optimizations on the Gate Network.} 
Our modeling assume that the gate network activates experts evenly, and many works focused on how to achieve this. 
% the consumption of load balance without performance degradation. 
For example,  
Lewis et al.~\cite{base-layer} proposed the BASE layer with token-to-expert allocation schema.
Zhou et al.\cite{expert-choice-route} proposed to allow experts to choose tokens.
HybridEP can integrate them.

\textbf{Optimizations on A2A Communication.} 
% We design HybridEP with the domain-based partition and expert compression, while the data compression algorithm and other A2A optimizations can be used from existing works~\cite{tutel, ZFP}. Specifically, 
To reduce A2A time, existing works focus on improving bandwidth utilization and reducing communication volume. For example, HetuMoE~\cite{hetumoe} proposed a hierarchical A2A algorithm to reduce communication rounds of inter-node communications; \cite{tutel, deepspeedmoe} proposed the 2D-hierarchical A2A algorithms to better utilize high-speed intra-node links; Zhou et al.~\cite{ZFP} used ZFP compression to reduce the A2A traffic.

% Some works aim to optimize the all-to-all communication algorithm to reduce the A2A time.
% Works to optimize A2A during MoE training includes optimizing bandwidth usage or reducing communication volume by data compression.
% HetuMoE~\cite{hetumoe} proposed a hierarchical A2A algorithm that reduces the communication rounds in inter-node communications.
% Recent works~\cite{tutel, deepspeedmoe} proposed the 2D-hierarchical A2A algorithms that try to better utilize intra-node bandwidth.
% In terms of data compression, Zhou et al.~\cite{ZFP} applied ZFP in MoE to reduce the A2A communication volume while preserving convergence.

\textbf{Optimizations on Comp. \& Comm. Overlap.} HybridEP combines prefetch and pipeline to fully overlap computation and communication, while existing studies try to optimize one of them as much as possible. For example, \cite{tutel, linamoe, pipemoe} try to find the optimal pipeline degree to fully overlap expert computation and A2A communication, while Janus~\cite{Janusmoe} tries to increase overlap time by pre-fetching experts. 
\section{Conclusion}
This paper presents \textbf{HybridEP}, a modeling-guided
framework that can efficiently scale EP under constrained bandwidth.
Through modeling, the appropriate strategy depends solely on the proportion of data and expert transmission traffic. 
Based on the modeling, we then introduce two techniques: domain-based partition and parameter-efficient migration to make our modeling practical, efficient, and scalable. Experiments suggest that HybridEP outperforms state-of-the-art MoE systems by up to 5.6$\times$.
On large scale simulation with over 1000 DCs under different bandwidths, HybridEP achieves up to 1.45$\times$ speedup compared to the original EP. 

% In this work, we proposed a new MoE training framework, named ACMoE, to accelerate distributed training on GPU clusters. 
% ACMoE alleviates the imbalanced load on heterogeneous bandwidth by transmitting experts. 
% First, ACMoE proposes a parameter-efficient transmission strategy using redundancy to achieve efficient expert transmission. 
% Second, ACMoE proposes domain-based transmission management, using structured expert domains to hierarchically manage multiple expert transmissions. 
% Finally, ACMoE proposes a switching-friendly pipeline to minimize switching overhead and achieve efficient computing.
% Experimental results show that ACMoE outperforms state-of-the-art MoE systems by around 1.23$\times$-2.31$\times$ on different configurations.

%%%%%%%%% -- BIB STYLE AND FILE -- %%%%%%%%
\bibliographystyle{IEEEtranS}
\bibliography{refs}
%%%%%%%%%%%%%%%%%%%%%%%%%%%%%%%%%%%%

\end{document}